\documentstyle{l-aa}

\def\etal{{\rm et al.\thinspace}}
\def\eg{{\it e.g.\ }}

\def\ie{{\it i.e.\ }}

\def\k{{\rm k}}

\def\hi{H{\sc i} \,}

\def\nH{\hbox{$N_{\rm H}$}}

\def\h50{\hbox{$h_{50}$\,}}
\def\tpow#1{$10^{#1}\,$}			
\def\spose#1{\hbox to 0pt{#1\hss}}
\def\ltsimm{\mathrel{\spose{\lower 3pt\hbox{$\sim$}}
	\raise 2.0pt\hbox{$<$}}}
\def\ltsim{$\mathrel{\spose{\lower 3pt\hbox{$\sim$}}
	\raise 2.0pt\hbox{$<$}}$}
\def\gtsimm{\mathrel{\spose{\lower 3pt\hbox{$\sim$}}
	\raise 2.0pt\hbox{$>$}}}
\def\gtsim{$\mathrel{\spose{\lower 3pt\hbox{$\sim$}}
	\raise 2.0pt\hbox{$>$}}$}

\def\fract#1/#2{\leavevmode\kern.1em                   
   \raise.5ex\hbox{\the\scriptfont0 #1}\kern-.1em
   /\kern-.15em\lower.25ex\hbox{\the\scriptfont0 #2}}
\def\deg{\hbox{$^\circ$}}
\def\arcm{\hbox{$^\prime$}}
\def\arcs{\arcm\hskip -0.1em\arcm}
\def\cm{{\rm\thinspace cm}}
\def\erg{{\rm\thinspace erg}}

\def\K{{\rm\thinspace K}}
\def\keV{{\rm\thinspace keV}}
\def\km{{\rm\thinspace km}}
\def\kpc{{\rm\thinspace kpc}}
\def\sol{\mbox{$_{\odot}$}} 
\def\Lsol{\hbox{$\thinspace L_{\odot}$}}

\def\Mpc{{\rm\thinspace Mpc}}
\def\Msol{\hbox{$\thinspace M_{\odot}$}}
\def\pc{{\rm\thinspace pc}}
\def\s{{\rm\thinspace s}}

\def\yr{{\rm\thinspace yr}}

\def\ps{\hbox{\s$^{-1}\,$}}
\def\pcc{\hbox{$\cm^{-3}\,$}}
\def\pyr{\hbox{$\yr^{-1}\,$}}

\def\pcm2{\hbox{$\cm^{-2}\,$}}
\def\ergpcm3ps{\hbox{$\erg\cm^{-3}\s^{-1}\,$}}
\def\ergps{\hbox{$\erg\s^{-1}\,$}}

\def\kmps{\hbox{$\km\s^{-1}\,$}}

\def\Lsolppc3{\hbox{$\Lsol\pc^{-3}\,$}}
\def\Msolppc3{\hbox{$\Msol\pc^{-3}\,$}}

\begin{document}

\thesaurus{11 (09.10.1; 
    11.09.1 M82; 
    11.19.3; 
    13.25.2)} 

\title{{\em ROSAT} Observations of the Galactic Wind in M82}
\author{D. K. Strickland \and T. J. Ponman \and I. R. Stevens}

\offprints{D. K. Strickland}

\institute{School of Physics and Space Research,
The University of Birmingham,\\
Edgbaston, Birmingham, B15 2TT, United Kingdom}

\date{Received ?; Accepted 22 July 1996}

\maketitle

\begin{abstract}
We present {\em ROSAT} PSPC and HRI observations of the galactic wind
from the starburst galaxy M82. \mbox{X-ray} emission from the wind is
detected to a distance of $\sim6\kpc$ from the plane of the galaxy.
Making use of the PSPC's mixture of good
spatial and spectral characteristics, we separate point source and
diffuse emission, and investigate the spectral variation of the diffuse
emission through the wind. The intrinsic X-ray luminosity of the
wind in the 0.1-2.4\keV\ band is found to be approximately
$1.9\times10^{40}$\ergps outside the immediate vicinity of the nucleus.
The temperature of
the diffuse emission is found to decrease weakly from $\sim0.6\keV$ to
$\sim0.4\keV$ along the minor axis, whilst the inferred gas density drops
as $z^{-0.5}$ and $z^{-0.8}$ along the northern and southern minor axes
respectively. We compare these results with those expected from two
simple models for the emission: Chevalier \& Clegg's adiabatically
expanding free wind and emission from shocked clouds in a wind, and find
that the emission cannot come from a free wind, but that shock heated
clouds could be the source of the emission.


\end{abstract}

\section{Introduction}
Only within the last two decades has it been realised that
the interplay between
vigorous star formation and the state of the interstellar medium (ISM)
can have profound
implications for the evolution of galaxies and their environments
(see for example Norman \& Ikeuchi 1989). Starbursts,
and in particular the galactic mass outflows or winds driven by thermalised
stellar winds from massive stars and supernovae, have implications
for systems of all sizes. Galactic winds may be responsible for the
destruction of dwarf galaxies (Dekel \& Silk 1986; Heckman \etal 1995),
enrichment of the ICM and IGM in clusters and groups and removal
of gas from merger remnants.

Heckman \etal (1993) provide a comprehensive review of the observational
data and theory of galactic winds.
Briefly, thermalised kinetic energy from stellar winds and supernovae from
massive stars in the starburst creates a hot ($\sim10^{8} \K$)
bubble in the ISM.
This expands, sweeping up ambient material into a dense
shell. Eventually the bubble breaks out of the disk of the galaxy along the
minor axis. The hot wind then escapes freely at several thousand kilometers
per second. The dense shell fragments due to Rayleigh-Taylor instabilities,
and is carried along by the wind at velocities of order hundreds of
kilometers per second. This or ambient clouds overrun by the wind is the
source of optical emission line filaments, and possibly the
soft \mbox{X-ray} emission.

The
archetypal starburst M82 presents possibly the best test case, given
its proximity (3.63 \Mpc, Freedman \etal 1994)
and the wealth of observational data available.
Its high infrared luminosity ($L_{IR}=4\times10^{10} \Lsol$,
Rieke \etal 1993), disturbed morphology, population of supernova remnants
(Muxlow \etal 1994) and
luminous young super star clusters (O'Connell \etal 1995) are all
signatures of a strong burst of star formation. The starburst was probably
caused by a close interaction with M82's nearby (projected distance
$\sim 40 \kpc$) neighbour M81 about $2\times10^{8 } \yr$ ago
(Cottrell 1977), and a tidal bridge of \hi
connects the two galaxies (Yun \etal 1993).

A set of emission line filaments along M82's minor axis
show velocities consistent with gas motions along the surface of a
cone at $v=600 \kmps$ (Axon \& Taylor 1978): cooler material swept
out of the galaxy by the much hotter wind.
The ${\rm H\alpha}$ emission defines an outflow that has a radius $\sim440\pc$
at a distance of $z\sim180\pc$ above the galactic plane (G\"{o}tz
\etal 1990), and is approximately cylindrical for $z<350\pc$. At larger $z$ the blowout
flares out to a cone with an opening angle $\theta\approx30\deg$ (see Fig.~5 in
McKeith \etal 1995). 
Within $1 \kpc$ of the nucleus the inferred electron density in the
filaments decreases with increasing $z$. McKeith \etal (1995) claim this
is consistent with a $\rho \propto z^{-2}$ model, as would be expected if
the ${\rm H\alpha}$ filaments were in pressure equilibrium with freely
expanding wind such as that proposed by Chevalier \& Clegg (1985). A
similar density decrease was also inferred by McCarthy \etal (1987).

Additional evidence for a galactic wind is the synchrotron emitting radio
halo, extended preferentially along the minor axis (Seaquist
\& Odegard 1991), due to relativistic
electrons from supernovae swept out from the starburst region
by the wind. This has a maximum extent comparable to
the \mbox{X-ray} emission (this paper). A steepening of the spectral index
interpreted as arising from
energy loss by Inverse Compton (IC) scattering
of the electrons off IR photons, allows an estimate of the speed with
which the electrons are being convected outwards,
assuming re-acceleration in shocks to be
negligible. Seaquist and Odegard (1991)
claim a conservative estimate of the wind
velocity, allowing for the uncertainties, lies
in the range 1000-3000 $\kmps$, similar
to that predicted from theory
(Chevalier \& Clegg 1985; Heckman \etal 1993).

Schaaf \etal's (1989) suggestion that \mbox{X-rays} produced by this IC
scattering could be the source of the observed \mbox{X-ray}
emission is argued
against by Seaquist \etal (1991) who predict $L_{\rm IC}=10^{38} \ergps$,
in contrast with the value we derive below of $2\times10^{40} \ergps$ in the {\em ROSAT}
band.

\mbox{X-ray} observations should provide a direct method of testing the
galactic wind paradigm, given that thermalised stellar wind and supernovae
ejecta is expected to have a temperature of $\sim10^{8} \K$.
Previous X-ray observations of M82 have suffered from poor sensitivity,
poor spectral resolution and to a lesser extent poor spatial
resolution. Watson \etal (1984) detected several very
luminous ($\sim10^{39} \ergps$) sources along with
diffuse emission using the {\em Einstein} HRI. The diffuse emission was seen
to extend out to $\sim3'$ ($\sim3\kpc$) to the south-east and $\sim2'$ to the
north-west. Spectral fits using the {\em Einstein}
IPC and MPC were inconclusive
in that they were unable to distinguish between a power law or a thermal
origin for the emission. Given that they were unable to separate the point
sources and the diffuse emission, this is not surprising.

A reanalysis of the {\em Einstein} data by Fabbiano (1988) did attempt to
separate the different components. The MPC (without any imaging
capability) fitted temperature of $6.8^{+5.7}_{-2.3} \keV$ would be
dominated
by the nuclear source and hence is not an estimate of the wind temperature.
The IPC gives $T\sim1.2 \keV$ for the emission within
$3'$ of the nucleus, compared to $\sim2.7 \keV$ for the
emission between 200-300$\arcs$.
The radial surface brightness in the IPC falls off approximately
as $r^{-3}$, consistent with the expectation for a free wind.

Schaaf \etal (1989) use an {\em EXOSAT} observation together with
the {\em Einstein}
data. The {\em EXOSAT} spectrum is consistent with either a power law or
a Raymond \& Smith plasma with temperature $9^{+9}_{-4} \keV$, again
without any separation of point source and diffuse components.

Although the extent of the emission seen within the {\em EXOSAT} and
{\em Einstein} observations compares well with the higher sensitivity
observation of {\em ROSAT} (Fig. 1), Tsuru \etal (1990) from observations
with Ginga, claim evidence for a very extended,
$100\kpc$ halo. Two north-south scans of a $5\deg$ region centred on M82
show excess flux to the north of M82, but not to the south. A spectral fit
to this emission is essentially unconstrained, with a temperature
in the range
1-11 \keV. Tsuru \etal argue that a single point source could not produce the
observed feature, as the position of the extra source is
inconsistent between the two scans. They concede this could be
due to two or more point sources, but estimate the chance
of finding two sources of suitable flux
in such a small region as 4 square degrees is $<5$\%. The dynamical age
of such a halo is $\sim10^{8} \yr$, hence this might be the remains
of a wind from a starburst  $\sim10^{8} \yr$ ago.

The {\em ROSAT} HRI observations of M82 (Bregman \etal 1995) show
three sources within the nuclear region of M82, although two of them have
very low S/N values above the strong and spatially varying wind emission.
A very bright source present in the {\em Einstein} data appears to have
faded away completely (Collura \etal 1994), although the main
nuclear source
is at a position consistent with the {\em Einstein} observation.

Bregman \etal (1995 )analyse the diffuse emission without the benefit of any
spectral
information. They conclude that the extended emission along the minor
axis is consistent with an outflow of gas with opening angle that decreases
with increasing radius
within $1.7'$ of the nucleus and at constant opening angle
at larger radii. They model the emission successfully by adiabatically
expanding gas of constant mass flux, and predict a decrease in
temperature of the gas with increasing radius.

\begin{figure*}
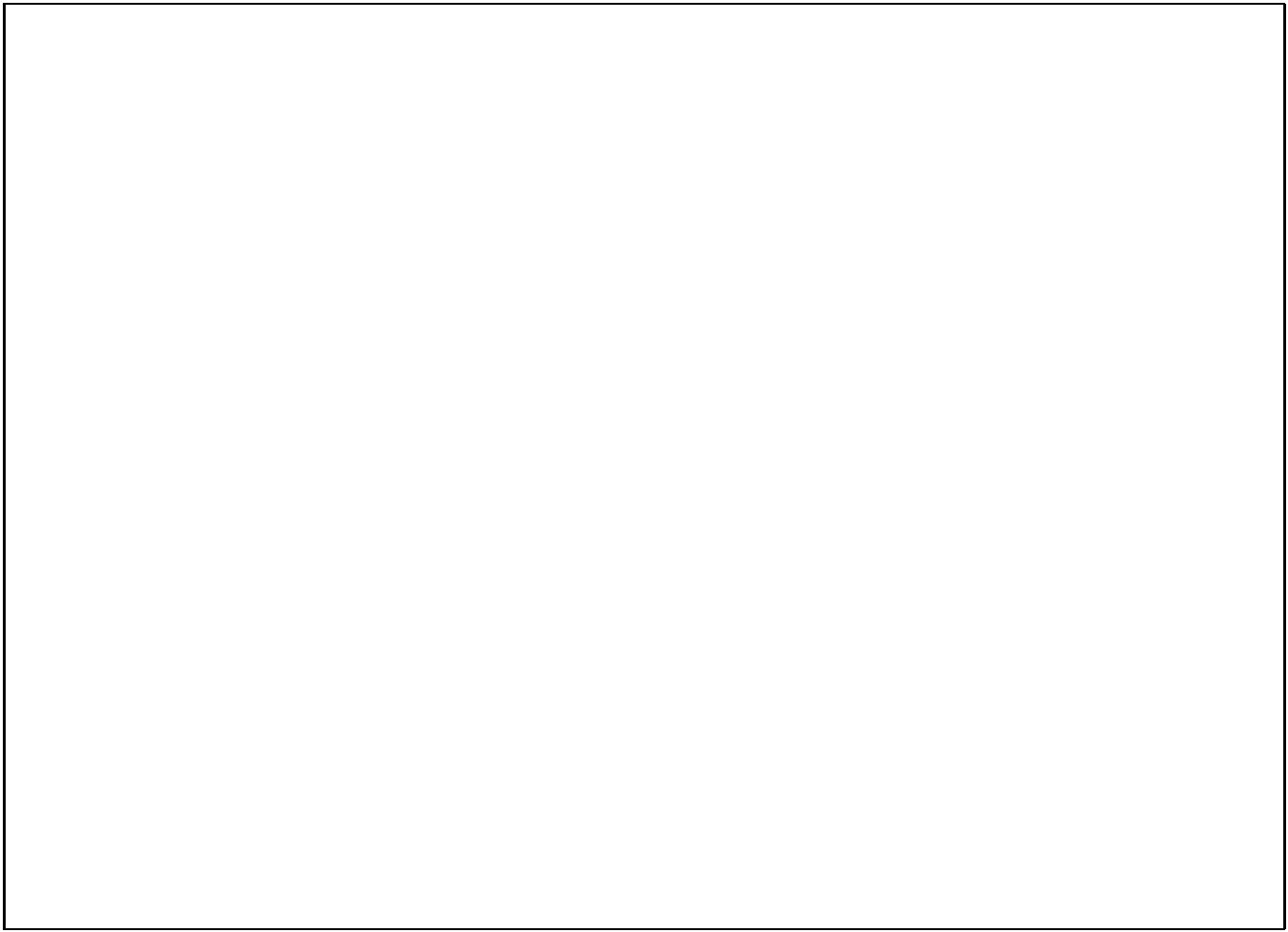

\picplace{13cm}
\label{fig:wind-grey}
\caption[]{Contours of X-ray emission ($0.1$--$2.4 \keV$) from the PSPC
overlaid on a digitised \mbox{sky-survey} optical image of M82. The
X-ray emission has been lightly smoothed with a Gaussian
of standard deviation
$\sigma=10\arcs$ to suppress noise. The contour levels increase
in factors of two from $2.88 \times
10^{-3}$ cts \ps arcmin$^{-2}$ ($\sim 3\sigma$ above the background).}
\end{figure*}

We report below, an analysis of the {\em ROSAT} PSPC and HRI
observations of this
\mbox{X-ray} emission. The PSPC's mixture of good spatial and spectral
capabilities compared to any other \mbox{X-ray} instrument, allow the
best determination yet of the properties of this emission. In particular,
we can separate point source and diffuse emission, and investigate
the variation of spectral properties as a function of distance from
the nucleus.
For the first time, we show that the diffuse emission is thermal in origin,
and obtain temperatures, emission measures, metallicities,
and, for an assumed geometry, electron densities, gas masses
and total energies. We compare our results with Chevalier \& Clegg's (1985)
analytical model of a galactic wind, and a simple model in which the
emission comes from shock heated clouds rather than the wind
itself.
Our results allow us to reject the possibility that the \mbox{X-ray}
emission comes from the wind itself, and show that it could be consistent with
emission from shock heated clouds.

\section{Data reduction}
\label{sec:red}
M82 was observed three times early in the {\em ROSAT} mission
(Tr\"umper 1984) by both the PSPC and the HRI (Table~\ref{tab:rosobs}).
Only the PSPC and the longer of the two HRI observations are used in
the present analysis. The use of the HRI's good spatial
resolution ($\sim 5\arcs$ FWHM) to complement the spectral
information available at moderate resolution
($27\arcs$ at 1 \keV) from the PSPC, is advantageous, especially with
regard to clarifying source confusion.
The data sets were obtained from the Leicester Data Archive (LEDAS)
and were analysed using the Starlink {\em ASTERIX} X-ray analysis system.

\begin{table*}
\caption[]{ROSAT observations of M82. Although the first HRI observation
rh600021 was ostensibly taken within a couple of days of the PSPC
observation, only $\sim 170$ seconds were taken in March 1991, the rest
being taken in early May 1991.}
\begin{flushleft}
\begin{tabular}{lllll}
\hline\noalign{\smallskip}
Instrument & Exposure (s) & ROR \# & P.I. & Start date \\
\noalign{\smallskip}
\hline\noalign{\smallskip}
HRI & 24613 & rh6200021 & Bregman & 25 03 1991 \\
PSPC & 26088 & rp600110 & Watson & 28 03 1991 \\
HRI & 9496 & rh600021ao1 & Bregman & 20 10 1992 \\
\noalign{\smallskip}
\hline
\end{tabular}
\end{flushleft}
\label{tab:rosobs}
\end{table*}

\subsection{Background subtraction of PSPC data}
\label{sec:bgsub}
The data were cleaned of periods of high background (both particle and Solar)
and poor pointing stability, leaving 21194 seconds of good data. A spectral
image (or {\em data cube}) was formed over a $0.3\deg \times 0.3\deg$
region centred on the $2.2 {\rm \mu m}$ nucleus, with
a pixel size of $5\arcs$ and 22
energy bins between channel numbers 11 and 230  (corresponding to roughly
0.11-2.3 keV). A model of the background was constructed using data
from an annulus $r = 0.15\deg-0.25\deg$, centred on M82, with contaminating
point sources removed.
The particle contribution to the background was estimated
using the master veto rate (Snowden \etal 1992) and the remainder
was corrected for energy-dependent vignetting, to
give a spatial-spectral model of the background covering the entire
field.
This model was then adjusted (by 3\%) as a result of an iterative process
of further source searching using the PSS programme (Allan 1995), 
removal of sources from the background annulus, and
rescaling of the background to achieve a drop to zero surface brightness
away from M82. The resulting background subtraction should be accurate to
2\%.


\subsection{HRI reduction}
\label{sec:hri-red}

The HRI's resolution and relative insensitivity to diffuse emission, make
it an ideal instrument for investigating point sources in the field,
clarifying the PSPC analysis. The data were binned into an image of size
$0.3 \deg \times 0.3 \deg$ centred on the $2.2 {\rm \mu m}$ nucleus, with
a $3 \arcs$ pixel size to exploit the HRI's superior spatial resolution.
Given the flatness of the HRI vignetting function within the region of
interest, no vignetting correction was applied. Source searching is
described below.

\begin{figure*}
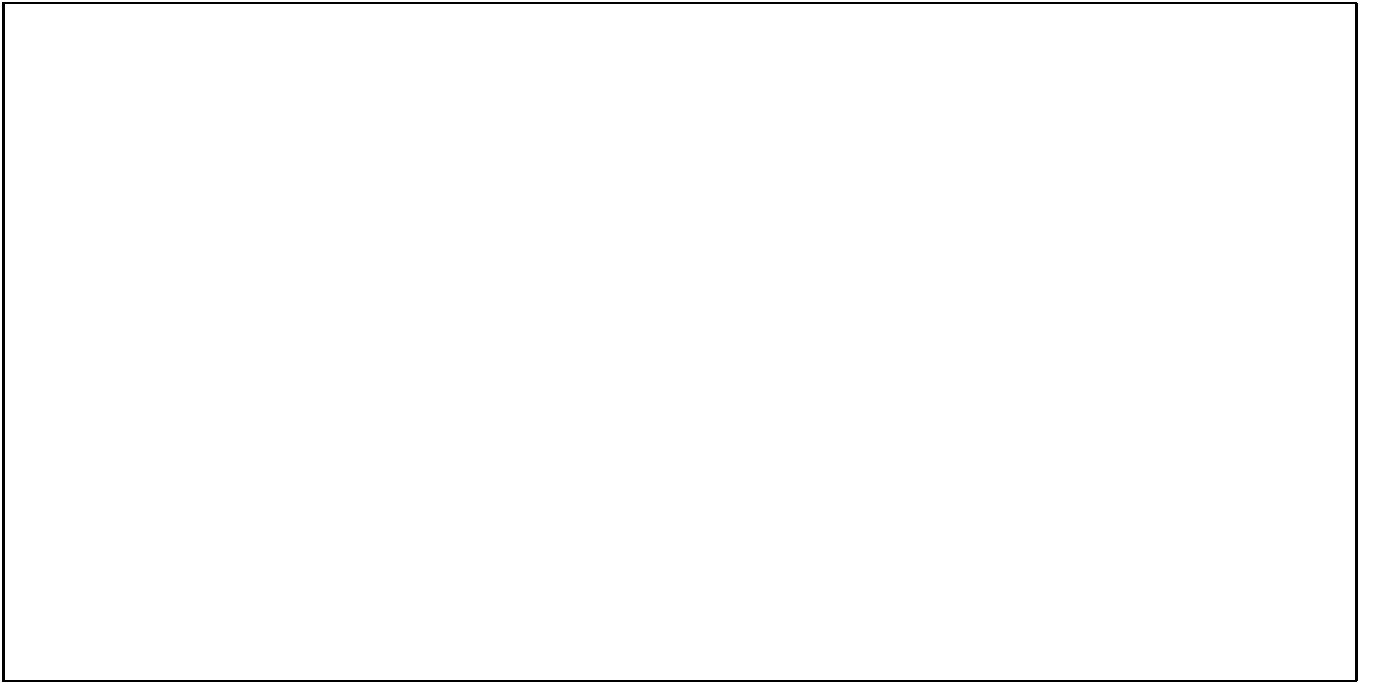

\picplace{9cm}
\label{fig:hri}
\caption[]{Background subtracted HRI image of M82. The data have been lightly
smoothed with a Gaussian of $\sigma=6\arcs$.
Contour levels begin at $4.0\times10^{-3}\, {\rm
counts} \ps$ arcmin$^{-2}$, an increase in factors of two.
HRI sources detected as described in the text are marked by
crosses, and are numbered as in Table 2.}
\end{figure*}

\begin{figure*}
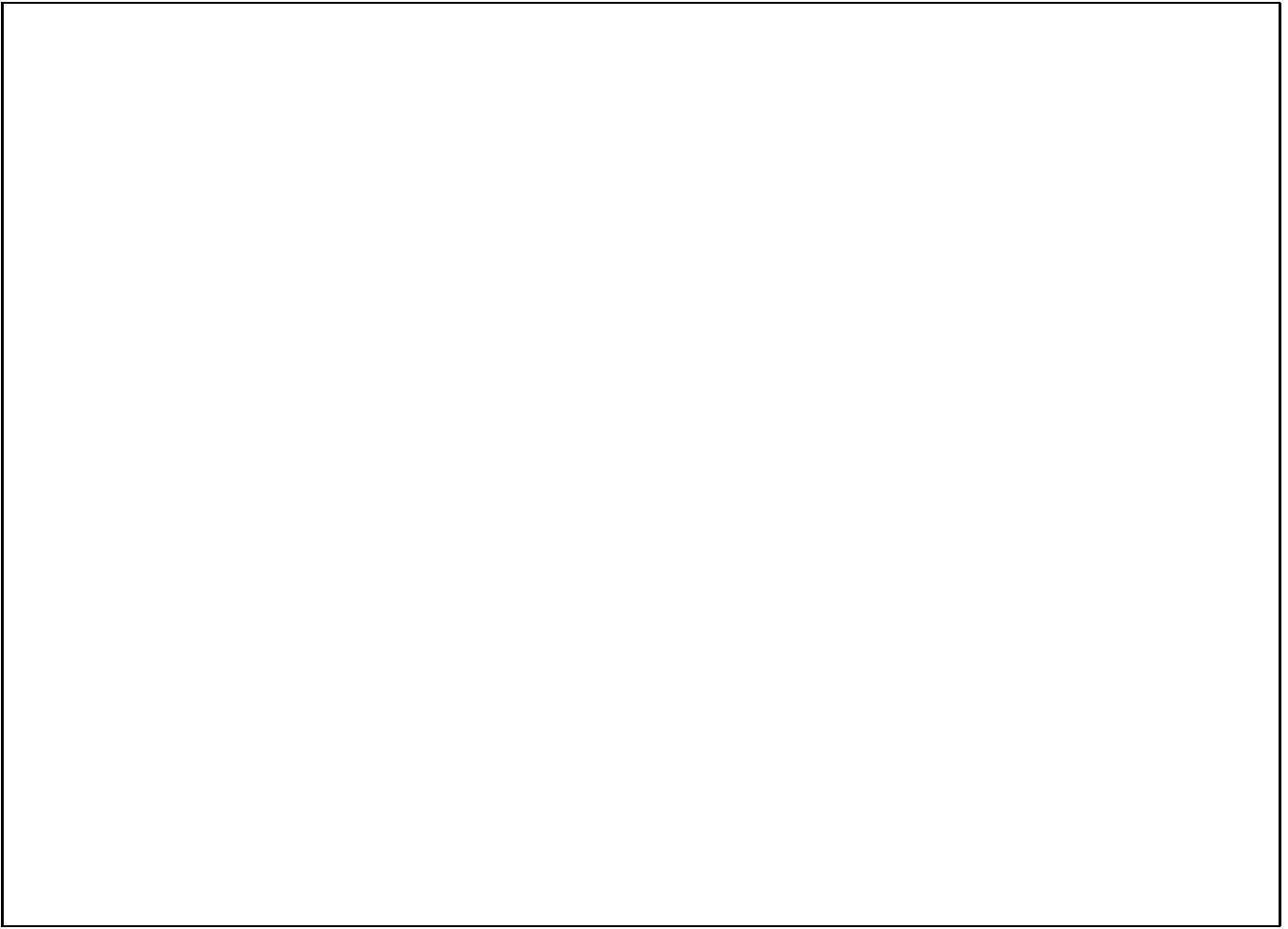

\picplace{13cm}
\caption[]{The wind regions from which spectra were collected, overlaid
onto an X-ray image. Sources (shown as circles and numbered as in Table
2) were removed from the
analysis. Contour levels are as in Fig. 1.}
\label{fig:wind-reg}
\end{figure*}

\section{Point sources}
\begin{table*}
\caption[]{PSPC and HRI detected sources. Positional errors are at  90\% confidence,
flux errors are 1 $\sigma$. Where a source is detected both in the PSPC and the HRI
the HRI determined position is given.}
\begin{flushleft}
\begin{tabular}{lllllll}
\noalign{\smallskip}
\hline
\noalign{\smallskip}
Source & RA & Dec. & Pos. Err. & Flux (PSPC) & Flux (HRI) & Identification \\
& (J2000.0) & (J2000.0) & (arcmin) & Counts & Counts & \\
\noalign{\smallskip}
\hline
\noalign{\smallskip}
1 & 09 54 20 & +69 46 42 & 0.10 & $ 41\pm{8} $ &  & \\
2 & 09 54 34 & +69 48 13 & 0.17 & $ 33\pm{8} $ &  & \\
3 & 09 55 07 & +69 43 40 & 0.11 & $ 40\pm{9} $ &  &
	 Wind enhancement? \\
4 & 09 55 15 & +69 36 19 & 0.04 & $ 39\pm{8} $ & $ 17\pm{7} $ & \\
5 & 09 55 16 & +69 47 41 & 0.06 & $ 120\pm{13} $ & $ 45\pm{9} $ & \\
6 & 09 55 33 & +69 44 47 & 0.10 & $ 62\pm{12} $ &  &
	Wind enhancement? \\
7  & 09 55 47 & +69 41 28 & 0.03 &  & $ 48\pm{13} $ & \\
8 & 09 55 51 & +69 40 48 & 0.01 & $ 8398\pm{108} $ & $ 1195\pm{48} $ &
 	M82 nuclear source \\
9 & 09 56 02 & +69 41 13 & 0.02 &  & $ 49\pm{11} $ & \\
10 & 09 56 18 & +69 49 02 & 0.04 &  & $ 18\pm{6} $ & \\
11 & 09 56 22 & +69 38 53 & 0.04 &  & $ 15\pm{6} $ & \\
12 & 09 56 43 & +69 38 05 & 0.17 & $ 37\pm{9} $ &  &
	Wind enhancement? \\
13 & 09 56 59 & +69 38 52 & 0.04 & $ 35\pm{8} $ &  $ 23\pm{7} $ &
	 QSO 0952+698 \\
14 & 09 57 00 & +69 34 21 & 0.11 & $ 57\pm{9} $ &  & \\
15 & 09 57 13 & +69 44 18 & 0.04 &  & $ 15\pm{6} $ & \\
16 & 09 57 23 & +69 35 36 & 0.03 & $ 67\pm{10} $ & $ 26\pm{8} $ & \\
\noalign{\smallskip}
\hline
\end{tabular}
\end{flushleft}
\label{tab:pspc-srcs}
\end{table*}

\begin{table*}
\caption[]{PSPC source properties. Error bounds are 1 $\sigma$.
The best-fit parameters for a Raymond-Smith
hot plasma model and a power law model are given.}
\begin{flushleft}
\begin{tabular}{lllllll}
\noalign{\smallskip}
\hline
\noalign{\smallskip}
Source & \multicolumn{3}{l}{Raymond-Smith}
 & \multicolumn{3}{l}{Power Law} \\
& $\nH$ & T & EM &  
$\nH$ & $\alpha$ & normalisation \\
& {\scriptsize (\tpow{21} $\pcm2$)} & 
	 {\scriptsize ($\keV$)} & {\scriptsize (\tpow{55} $\cm^{3}$ / $10 \kpc^{2}$)}
& {\scriptsize (\tpow{21} $\pcm2$)} && 
	{\scriptsize (\tpow{-5} \mbox{photons} $\pcm2 \ps \keV^{-1}$)} \\
\noalign{\smallskip}
\hline
\noalign{\smallskip}
1	& $ 0.28^{+0.28}_{-0.17} $ & $ 1.78^{+20}_{-0.47} $ 
		& $ 5.4^{+5.7}_{-2.0} $
	& $ 0.48^{+0.44}_{-0.31} $ & $ 2.3^{+1.0}_{-0.9} $
		& $ 0.8^{+0.2}_{-0.2} $ \\
2 	& $ 0.35^{+0.73}_{-0.35} $ & $ 0.88^{+1.32}_{-0.40} $ 
		& $ 4.9^{+9.8}_{-4.6} $
	& $ 0.70^{+1.16}_{-0.47}  $ & $ 2.7^{+1.3}_{-1.2} $ 
		& $ 0.7^{+0.4}_{-0.2}  $ \\
3 	&  $ 4.2^{+7.6}_{-4.2} $ & $ 0.23^{+0.36}_{-0.18} $ 
		& $ 48.7^{+26580}_{-48.7} $
	& $ 7.5^{+0.9}_{-2.5} $ & $ 8.0\pm{1.6} $ 
		& $ 12.4^{+4.6}_{-6.6} $ \\
4	& $ 0.00^{+0.06}_{-0.00} $ & $ 0.57^{+0.18}_{-0.17} $ 
		& $ 2.0^{+1.7}_{-0.7} $
	& $ (0.19^{+0.33}_{-0.19}) $ & $ (2.2\pm{1.0}) $ 
		& $ (0.7^{+0.1}_{-0.2}) $ \\
5	& $ 0.79^{+1.26}_{-0.28} $ & $ 0.65^{+0.14}_{-0.34} $ 
		& $ 32.1^{+59.1}_{-16.1} $
	& $ 8.5^{+0.7}_{-3.2} $ & $ 8.0\pm{2.2} $ 
		& $ 31.7^{+5.4}_{-14.8} $ \\
6	& $ 0.87^{+0.30}_{-0.20} $ & $ 0.36^{+0.08}_{-0.06} $ 
		& $ 100.3^{+97.8}_{-48.6} $
	& $ 1.77^{+0.79}_{-0.36} $ & $ 4.6^{+0.7}_{-0.5} $ 
		& $ 48.0^{+12.4}_{-6.3} $ \\
8	& $ (3.17^{+0.24}_{-0.17}) $ & $ (2.39^{+0.48}_{-0.43}) $ 
		& $ (3770^{+30}_{-20}) $
	& $ 3.96^{+0.29}_{-0.28} $ & $ 2.17^{+0.14}_{-0.13} $ 
		& $ 717^{+56}_{-44} $ \\
12	& $ 0.9^{+3.4}_{-0.4} $ & $ 0.43^{+0.24}_{-0.30} $ 
		& $ 36^{+144}_{-29} $
	& $ 5^{+6}_{-1} $ & $ 6.3^{+4.2}_{-2.2} $ 
		& $ 4.3^{+8.8}_{-4.3} $ \\
13	& $ 2^{+7}_{-2} $ & $ 0.7^{+3.5}_{-0.2} $ 
		& $ 24^{+409}_{-24} $
	& $ 4^{+7}_{-4} $ & $ 3.6^{+3.7}_{-2.2} $ 
		& $ 2.8^{+10.9}_{-2.8} $ \\
14	& $ 0.10^{+0.10}_{-0.08} $ & $ 1.4^{+3.6}_{-0.6} $ 
		& $ 6.1^{+4.4}_{-2.0} $
	& $ 0.21^{+0.21}_{-0.15} $ & $ 2.2\pm{0.7} $ 
		& $ 0.8\pm{0.2} $ \\
16	& $ 0.24^{+0.18}_{-0.14} $ & $ 2.9^{+21.2}_{-0.5} $ 
		& $ 9.0^{+4.6}_{-2.3} $
	& $ 0.32^{+0.28}_{-0.22} $ & $ 1.7\pm{0.7} $ 
		& $ 1.4^{+0.3}_{-0.2} $ \\
\noalign{\smallskip}
\hline
\end{tabular}
\end{flushleft}
\label{tab:psrcs-fits}
\end{table*}

\subsection{Source searching in the presence of diffuse emission}
\label{sec:srcs}
For M82, source searching must take into account the presence of a spatially
variable high surface brightness diffuse background due to the wind.
This affects
both the PSPC and the HRI. Failure to incorporate this additional background
leads to the detection of large number of low significance
sources within the diffuse emission, whilst simply increasing the
threshold significance leads to the non-detection of what
are clearly real sources in regions free of diffuse emission.

As a result of this we employ an iterative procedure which starts
by using a smoothed image (including all sources and diffuse emission)
as the background for the source searching procedure. Sources detected at
$\gtsimm5\sigma$ are then excised from the image, out to the
$\sim80\%$ enclosed energy radius ($7\arcs$ in the HRI, $26\arcs$ in
the PSPC), and the resulting holes interpolated over.
This dataset is then smoothed to provide a second approximation to the
background. The 80\% radius, to which sources are removed,
is a compromise between removing so large a radius that the
interpolation is unreliable, and too small a
radius, leaving significant source contamination in the background.
The above procedure of source searching followed by background estimation,
was repeated until there was no further change in the list of detected sources.
In practice this required three cycles.
The results of the method depend on the smoothing scale employed
when estimating the background,
and $\sigma=18\arcs$ and $60\arcs$ were found to give best results
for the HRI and PSPC, respectively.
The final combined sourcelist is shown in Table~\ref{tab:pspc-srcs}.

\subsection{Source spectra}
Individual exposure-corrected spectra
centred on the positions given in Table~\ref{tab:pspc-srcs} were obtained
for each PSPC source from the data cube within a radius corresponding
to a 95\% enclosed energy fraction at an energy of $0.5 \keV$ (an
appropriate energy for QSO's which form the majority of background
sources). \mbox{Raymond \& Smith} (1977) and power law models were fitted
to these spectra. Standard $\chi^{2}$ fitting is inappropriate,
due to the low numbers of counts per bin; we therefore used a
\mbox{maximum-likelihood} fit. For each source,
the spectrum predicted from the spectral model is added to an
estimated background spectrum derived from the background
model cube discussed in Sect.~\ref{sec:bgsub}. The resulting total source
plus background spectrum is fitted to the observed spectral data using the
Cash C-statistic (Cash 1979).

Table~\ref{tab:psrcs-fits} gives the results of the spectral fits.
As maximum likelihood does not provide
a goodness-of-fit measure akin to $\chi^{2}$, it is difficult to assess
how good the fits are, except by visual inspection and comparing the
fitted parameters with typical values for QSO's and stars. Source 8 (the
nucleus) is strong enough that significant systematic discrepancies
between the data and the fitted models are apparent, and is dealt
with separately in Sect.~\ref{sec:nucleus}.

\subsection{Comparison between PSPC and HRI sources}
\label{sec:hrisrc}
Given the presence of several possible point sources within or in close
proximity to the diffuse emission, the use of the HRI can
clarify whether these are true point sources or not. As discussed in
Sect.~\ref{sec:hri-red} the presence of strong diffuse emission
complicates source searching, which may lead to the detection of
spurious sources within the diffuse emission.

Only five sources were detected independently in both the
PSPC and the HRI. We have searched for counterparts to these objects
at other wavelengths in the SIMBAD database. One (Source 8) is
M82's nuclear source,
another (Source 13) a QSO. The other three do not have any counterparts.

For the remaining PSPC sources not detected in the HRI, we found $3\sigma$ upper
limits for the HRI flux due to any point source within a region defined by the
PSPC positional uncertainty. These, together with predicted HRI count rates
from the spectral fits to the PSPC sources, are given in
Table~\ref{tab:hri_uplim}.  There is little difference between the predicted
fluxes for the power law and \mbox{Raymond \& Smith} models.  For the sources
detected in both PSPC and HRI, the observed HRI count rates agree well with
those predicted, except for the nucleus, which is discussed below. In most
other cases the upper limits are greater than (i.e. consistent with) the
predicted count rates. For the two possible sources within the northern wind,
sources 3 \& 6, the predicted count rates are higher than the HRI $3\sigma$
upper limits. This could mean that these are not true point sources, merely
bright lumps in the wind.
Note, however, that source 6 appears to be significantly cooler than the
temperature of the wind emission in region n7, where it is centred.
Alternatively, the predicted count rate could be
overestimating the flux, due to the addition of diffuse flux along with real
source flux. Variability is another possible explanation for the HRI
non-detection.

\subsection{The nucleus}
\label{sec:nucleus}
As the HRI observations have shown (Collura \etal 1994; Bregman \etal
1995), the nucleus of M82 contains a very luminous, variable \mbox{X-ray}
source, along with non-uniform diffuse emission. It is little surprise that the single
component fits to the nuclear source in the PSPC data (source 8 in
Table~\ref{tab:pspc-srcs}) are of poor quality.
Given the large number of counts in the spectrum ($11033\pm249$) we
can use standard $\chi^{2}$ fitting.
The power law from Table~\ref{tab:pspc-srcs} has a reduced $\chi^{2}$ of 7 with
19 degrees of freedom.
An additional problem is the finite radius of $34\arcs$ within
which the spectrum was extracted. Although
the 95\% enclosed energy radius at $0.5 \keV$ is good for almost all sources,
the brightness of the nucleus, coupled with its hardness, means that
significant flux is scattered outside this radius.  At energies above
$\sim1.75 \keV$ this radius only encloses $\sim70\%$ of the flux.
Given the large number of photons involved, this is likely to have a
significant effect on any fit.

Fitting the data from within a larger ($r=0.02\deg$, giving $\sim90$\% enclosed
energy at $1.75\keV$, $17473\pm415$ counts) 
radius from the nucleus, we achieve a best-fit reduced
$\chi^{2}$ of  2.08 with 15 degrees of freedom (see Table~\ref{tab:nuc}) using
a two component soft \mbox{Raymond \& Smith} plus harder bremsstrahlung model.
The fitted temperature of the hard component is outside {\em ROSAT}'s energy
range, and so should only be interpreted as being being significantly hotter
than 6.2 \keV.  The spectrum can also be fitted by a two component power law
plus \mbox{Raymond \& Smith} model, however the fit requires the power law to
have a {\it lower} column than the hot plasma component. This is not physically
sensible if the hard component represents nuclear emission, so we rejected this
model in favour of the \mbox{Raymond \& Smith} plus bremsstrahlung model.

\begin{table}
\caption[]{HRI count rates and upper limits for the PSPC detected sources.
The predicted HRI count rate was calculated as described in
the text. Only five PSPC sources are directly detected in the HRI, the
remaining sources have $3\sigma$ upper limits quoted. The range in
predicted flux corresponds to using either the power law or Raymond-Smith
fit.}
\begin{flushleft}
\begin{tabular}{lll}
\hline\noalign{\smallskip}
Source & Predicted flux & Observed Flux \\
& ($10^{-4}$ cts $\ps$) & ($10^{-4}$ cts $\ps$) \\
&& \\
\noalign{\smallskip}
\hline\noalign{\smallskip}
1 & 5.9--6.3 & $<$ 9.5 \\
2 & 4.9--5.4 & $<$ 9.4 \\
3 & 12.8--13.4 & $<$ 10.2 \\
4 & 6.1--6.9 & $6.6\pm2.6$ \\
5 & 21.8--24.3 & $18.8\pm3.8$ \\
6 & 25.8--29.5 & $<$ 12.5 \\
8 & 1857--1996 & $500\pm20$ \\
12 & 8.6--8.8 & $<$ 10.8 \\
13 & 7.1--7.2 & $9.3\pm3.0$ \\
14 & 7.9--8.0 & $<$ 9.7 \\
16 & 10.9-11.3 & $11.0\pm3.0$ \\
\noalign{\smallskip}
\hline
\end{tabular}
\end{flushleft}
\label{tab:hri_uplim}
\end{table}

\begin{table}
\caption[]{Best-fit parameters (reduced $\chi^{2}=2.08$)
for the nuclear source. Luminosities are quoted in the {\em ROSAT}
band ($0.1-2.4 \keV$), for a distance of $3.63 \Mpc$
(Freedman \etal 1994) to M82.
The luminosity escaping M82 $L_{X-esc}$ is corrected for absorption
in our own galaxy. The intrinsic source luminosities $L_{X-in}$
are corrected to zero absorption. $^{\dagger}$ This is the $1\sigma$
lower bound. The temperature is unbounded above this value.}
\begin{flushleft}
\begin{tabular}{lll}
\hline\noalign{\smallskip}
Parameter & Hard component& Soft component \\
& {\scriptsize bremsstrahlung} & {\scriptsize Raymond \& Smith} \\
\noalign{\smallskip}
\hline\noalign{\smallskip}
$\nH$ {\scriptsize($10^{21} \pcm2$)}
    & $5.76^{+1.49}_{-1.17}$ & $0.92^{+0.08}_{-0.12}$ \\
EM {\scriptsize(\tpow{57} cm$^{3} / 10 \kpc^{2}$)}
    & $36.7^{+2.6}_{-3.4}$ & $8.3^{+2.3}_{-2.5}$ \\
T {\scriptsize(\keV)}
    & $>6.2^{\dagger}$ & $0.76^{+0.02}_{-0.03}$ \\
Metallicity {\scriptsize($Z_{\odot}$)}
    & - & $0.30^{+0.09}_{-0.05}$ \\
$L_{X-esc}$ {\scriptsize (\ergps)}
    & $1.3\times10^{40}$ & $7.8\times10^{39}$\\
$L_{X-in}$ {\scriptsize (\ergps)}
    & $3.5\times10^{40}$ & $1.2\times10^{40}$ \\
\noalign{\smallskip}
\hline
\end{tabular}
\end{flushleft}
\label{tab:nuc}
\end{table}

The predicted HRI count rate for the hard bremsstrahlung component
derived above is $\sim0.15$ cts $\ps$, compared to the observed value of
$\sim0.05$ cts $\ps$. The thin, hot component of the wind is predicted to
provide very little emission in the {\em ROSAT} band (Suchkov \etal
1994), and there are no other strong point sources seen by the HRI. Since
the HRI source associated with the nucleus is known to be variable
(Collura \etal 1994) by a factor of several within the HRI observations,
the difference between PSPC and HRI fluxes is most likely due to such
variability.

The first HRI observation (rh600021, see Table~\ref{tab:rosobs}), is divided
into two blocks which are interleaved with the PSPC observation.  The short
($\sim170 \s$) initial HRI pointing showed the source intensity at a level
$\sim\frac{1}{3}$ that of the remaining HRI data, taken $\sim40$ days later.
The first block of PSPC data falls in this 40 day gap, with the remainder
commencing $\sim200$ days later.  We have searched for variation of the nuclear
source within the PSPC observation, which is broken into three main blocks
separated by long gaps, but observe no significant variation between the blocks.
In the second HRI observation, taken a year later, the nuclear count rate
decreases steadily over a period of six days from the previous HRI level to
about half that (Collura \etal 1994).

The intrinsic nuclear point source luminosity in the {\em ROSAT} band of $\sim3.5\times10^{40}
\ergps$ (Table~\ref{tab:nuc}) is significantly higher than the {\em ROSAT} HRI
estimate of $\sim6.3\times10^{39} \ergps$. This estimate assumed a
Raymond-Smith model with $T=3 \keV$ and $\log \nH=21.5$. With the higher
temperature and absorption column inferred from the PSPC spectral fit, the
HRI luminosity would increase, although the inferred PSPC luminosity
remains several times higher. This PSPC luminosity corresponds to the
Eddington luminosity for a $\sim250 \Msol$ object. The position of this
source corresponds (to within {\em ROSAT} pointing accuracy) to the
position of a strong $6 \cm$ radio source (41.5+597) which, on the basis
of a 100\% drop in flux within a year, is unlikely to be a supernova
remnant (Muxlow \etal 1994). A high surface brightness complex is seen
in the optical (region E in O'Connell \etal 1995) at this position. This
is an unusual object deserving more study.

\section{Wind properties}
\label{sec:wind-prop}

\begin{figure*}
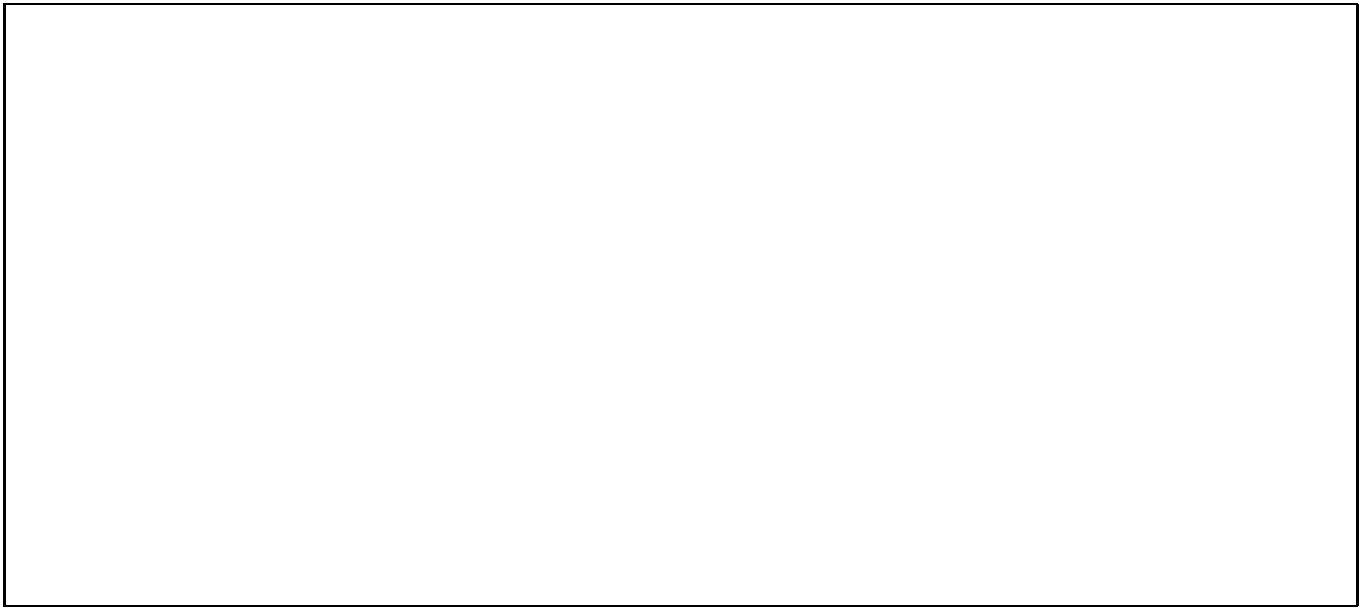

\picplace{8cm}
\caption{Azimuthal profile of the emission within $0.1\deg$ of
the nuclear source. Sources other than the nucleus have been masked out.
The northern minor axis is at an azimuth of $65\deg$, the southern minor
axis at $245\deg$.}
\label{fig:azimuth}
\end{figure*}

\subsection{Analysing the wind}
\label{sec:an_wind}
As our aim is to derive spatially resolved plasma parameters (temperature,
density, metallicity and absorbing column) it is necessary to assume
a geometry for the emission. This will dictate the regions from which
spectra are taken and the volumes used in deriving the density of the
emitting plasma. Given the lack of symmetry of the diffuse
emission (henceforth called the wind, bearing in mind alternative
explanations of its origin as wind-shocked ambient
material or even a hydrostatic halo as discussed in
Sect.~\ref{sec:discuss}) apparent in Fig.~\ref{fig:wind-grey}, it is not
obvious what geometry to choose.  Within the wind paradigm, a conical
(\eg Suchkov \etal 1994) to cylindrical outflow
(\eg Tomisaka \& Ikeuchi 1988; Tomisaka \& Bregman 1993) along
the galaxy's minor axis is expected, and if the emission arises
from \mbox{wind-shocked} material a similar geometry would apply.

The azimuthal profile (Fig.~\ref{fig:azimuth}) of the PSPC data about the
centre of the galaxy can be used to explore the geometry of the diffuse
emission.  A biconical outflow would result in a sharp drop in the azimuthal
brightness profile at the angles corresponding to the
edges of the cone.  If the cone were actually limb
brightened (as suggested in some models), then a bimodal structure would be
seen in the azimuthal profile of
each outflow. In practice, the profile varies quite smoothly with
azimuth in both hemispheres, and no suggestion of a limb brightening is
apparent. It appears that a conical geometry is a poor representation.

Inspection of the surface brightness shows the Northern wind
to be reasonably well described as a cylinder of radius $r = 0.05 \deg$.
We therefore adopt a cylindrical geometry for the bulk of our analysis.
For consistency, we apply the same geometry to the southern wind,
although it is clear from Fig.~3 that this is a poorer approximation,
which will overestimate the emitting volume.

A set of spectra along the northern and southern winds were
accordingly extracted from a series of
rectangular regions of width $0.1\deg$ and height $h$ along the
minor axis (Fig. 3).  A compromise must be made between
large $h$ (collecting a larger number of photons in the spectrum) and
small $h$ (giving good spatial resolution along the
wind). A value of $h=0.01 \deg$  was chosen, giving 9 regions
along the wind while still having sufficient photons to give reasonable
constraints on the spectra for all but the outermost regions. Corresponding 
background spectra were formed from the \mbox{background-model} cube
for each of the wind region to allow maximum likelihood fitting.

\begin{figure}
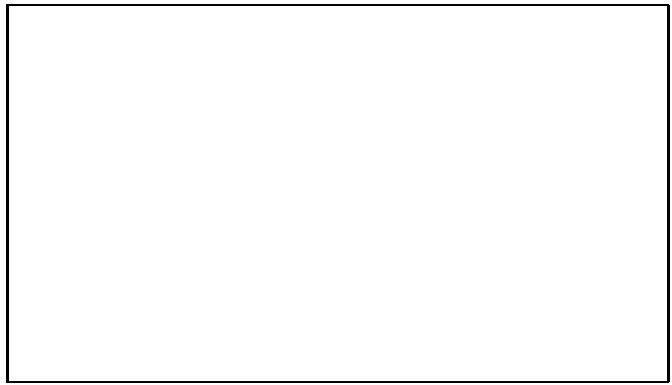

\picplace{5cm}
\caption[]{Cylindrical and conical geometries assumed for the spectral
analysis.}
\label{fig:cone}
\end{figure}

Since it is likely that there will in practice be some degree
of divergence of the outflows (though as we will show, the X-ray emission
is almost certainly not coming from the wind fluid itself), we have
investigated the effects of this by
performing an identical analysis using two truncated cones.
These truncated cones have a radius in the galactic plane $r_{\rm
base}=0.03\deg$, and an opening angle of $50\deg$. The height $h$ is
again $0.01\deg$. The cylindrical and diverging geometries are
compared in Fig.~\ref{fig:cone}.

As discussed in Sect.~\ref{sec:hrisrc}, several point-sources were
detected within or in close proximity to the diffuse wind emission.
While it is possible
that these represent regions of enhanced diffuse emission rather than truly
independent sources, they were masked out of the wind regions to
prevent any possible contamination of the wind spectra by foreign
flux. The spectra were then corrected for \mbox{dead-time} and
exposure corrected.

Raymond \& Smith (1977) hot-plasma models were then
fitted to the strip spectra using maximum likelihood, initially allowing
all parameters to optimise. It was found that
the metallicities consistently fitted low, $0.00$--$0.07 Z_{\odot}$.
In view of the present uncertainties in the accuracy of {\em ROSAT}
metallicities (see Bauer \& Bregman 1996), and the expectation that the
metallicity should not vary greatly through the wind, we refitted all the
spectra with the metal abundance fixed at $0.05 Z_{\odot}$, which reduced
the scatter in the other fit parameters. All results for the wind quoted
below are derived from these $Z=0.05 Z_{\odot}$ fits.

Such a low metallicity, whilst surprising, is supported by the results of
recent ASCA observations (Ptak \etal 1996; Tsuru 1996). ASCA has the
spectral resolution to clearly distinguish the iron L complex, which is
the strongest metallicity indicator for plasmas of this temperature, and
the implied iron abundances in the soft spectral component is 0.04--0.05
$Z_{\odot}$ (with a typical error of $\approx 0.02$), in good agreement
with our results.

\begin{figure*}
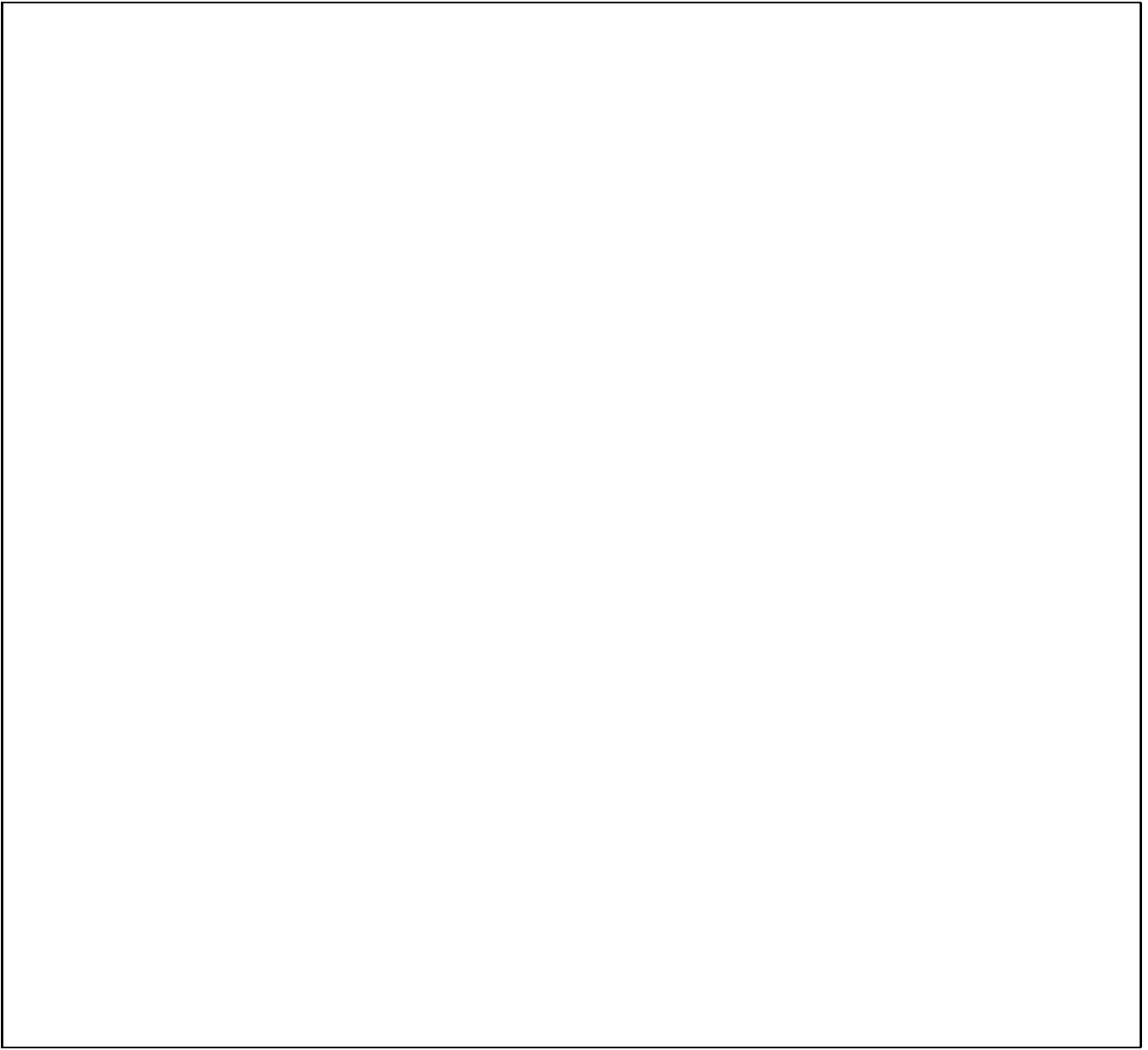

\picplace{16.5cm}
\caption{Error ellipses for metallicity against emission measure at 68\% confidence
for two interesting parameters, for all the wind regions.
Regions n7 and n8 are clearly peculiar as discussed in
Sect.~4.5. Region n5 is poorly constrained.}
\label{fig:zem}
\end{figure*}

Since emission lines are so strong in the {\em ROSAT} energy band for
$T\sim 0.5$\keV plasma, metallicity trades off against emission measure
in fitted spectra. This can be clearly seen in Fig.~\ref{fig:zem}, which
shows the error ellipses for 68\% confidence for {\em two} interesting
parameters (optimising temperature and absorbing column) for all of the
regions used in the analysis. Hence any error in our assumed metallicity
will lead to a corresponding error in derived emission measure, and hence
in inferred gas density. As can be seen, from Fig.~\ref{fig:zem}, the
error envelopes generally fall below $Z=0.2 Z_{\odot}$. Hence, taking
$Z=0.2 Z_{\odot}$ as the highest plausible value for metallicity (though
this falls well outside the ASCA errors), our emission measures would be
overestimated by a factor $\sim4$, and hence the densities would be a
factor $\sim2$ too high. Such an error is quite modest compared to those
introduced by the unknown filling factor and uncertain geometry of the
emitting material.

\begin{figure*}
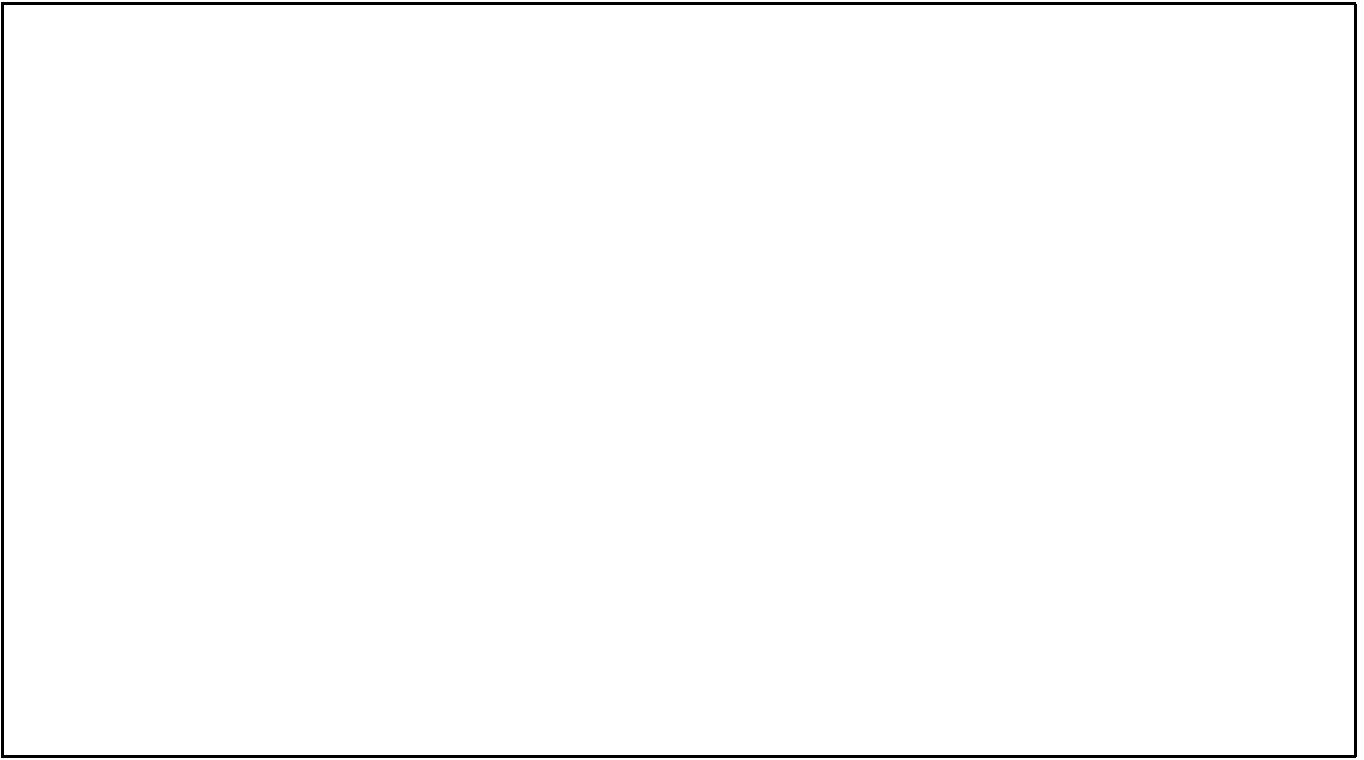

\picplace{10cm}
\caption[]{Spectral fits to two for two of the wind regions, n2
(left) and n6 (right), representative
of the range in quality of the spectra obtained.
Normalised background-subtracted
spectra are shown overlaid with power law (dashed line) and contamination
corrected Raymond \& Smith (solid line) best fits. For the
inner regions (such as n2) the
power law is clearly a poorer fit than the Raymond \& Smith model. }
\label{fig:fits}
\end{figure*}

As previous X-ray observations of M82 have been unable to determine
whether a hot plasma or a power law gives a better fit, we also fitted power
law spectra to the data. These were found to give significantly poorer
fits than the hot plasma fits (see \eg Fig.~\ref{fig:fits})
for all but the outer regions, where the statistics were too poor to tell.
Although maximum likelihood gives no absolute goodness of fit measure,
the relative likelihood between two fits to the same data can be
derived from the Cash
C-statistic. From Cash (1979), the relative probabilities $P_{\rm
2}/P_{\rm 1}=\exp{(\Delta C_{\rm 12}/ 2)}$,
where $\Delta C_{\rm 12}=C_{\rm
1} - C_{\rm 2}$ is the change in Cash statistic between the two fits.
For the inner wind, contamination-corrected Raymond \& Smith fits are
clearly superior, \eg for region n2, the hot plasma fit
is $\sim400$ times more probable than the best-fit power law.

\subsection{Nuclear contamination of the wind}
\label{sec:contam}
Given the presence of an extremely luminous hard point source
(nearly a third of all counts
detected from M82 with the PSPC are within a PSF sized
region of $r\sim 30\arcs$)
at the centre of M82, and the increasing size of the PSPC PSF
at higher energies, one expects some contamination of the wind emission in the
inner strips by photons from the nuclear source.

We can roughly assess the level of contamination by asking what fraction
of the flux within the two innermost wind regions (n1 and s1) is due to
scattered nuclear flux. If we assume all the flux within a $r=0.02\deg$
region centred on the nuclear source were due to a point source, then
the fraction of this flux scattered into wind region n1 is 15\% of
the total flux observed in n1. For the innermost southern region, s1, the value is
12\%. These are overestimates, as only $\sim\frac{1}{2}$ of the
flux within $r=0.02\deg$ is due to the point source.


We allow for nuclear contamination by using
two-component models for the wind regions: a soft
Raymond \& Smith plasma for the wind, and a harder bremsstrahlung
component for the nuclear contamination. The bremsstrahlung component
in each strip was fixed: the absorbing column and temperature
taking the values derived from the nuclear fit discussed
in Sect.~\ref{sec:nucleus},  and the contaminating flux being
estimated from the (energy dependent) PSPC point spread function.

\subsection{X-ray morphology}

\begin{figure*}
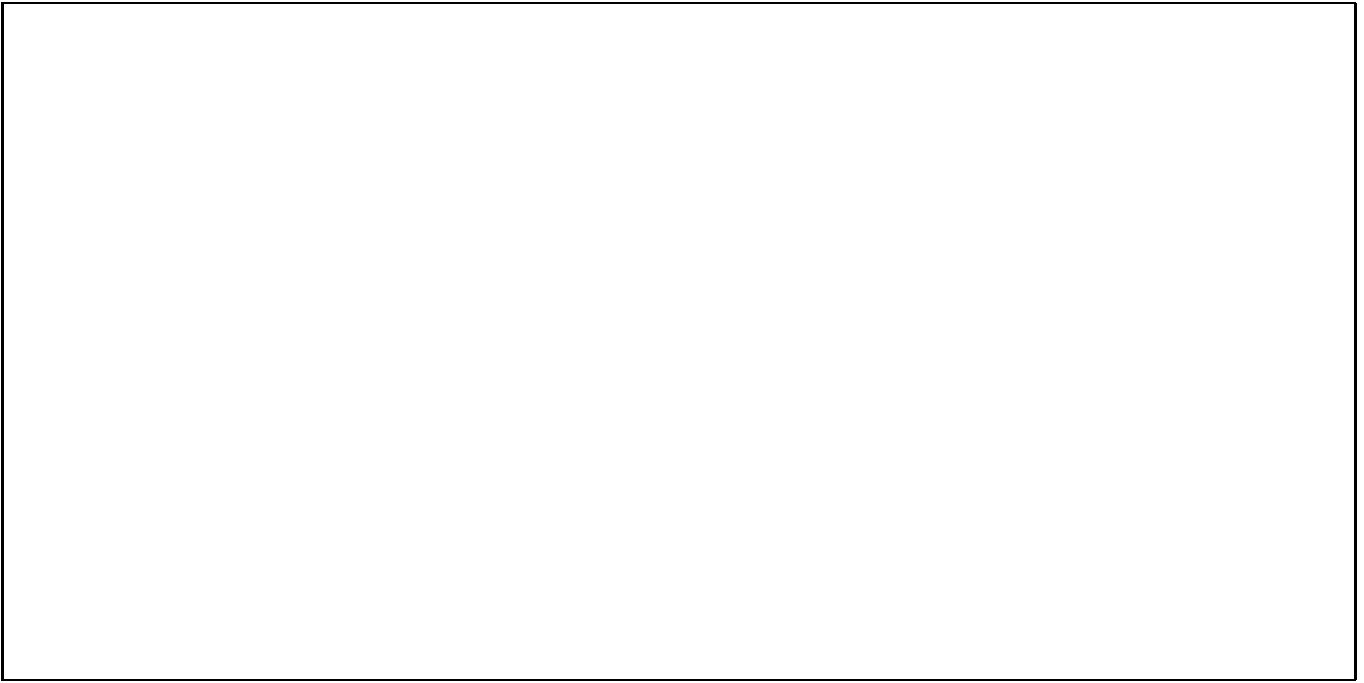

\picplace{9cm}
\caption{Surface brightness along the minor axis
in a slice $0.1\deg$ wide.
Diamonds represent the emission to the south, crosses the norther data.
In each case the line represents the data with sources
(other than the nuclear source)
excluded. The south is brighter than the north within $\sim2\kpc$
in both the \mbox{X-ray} and the radio (Seaquist \& Odegard 1991).
Diffuse emission extends out to $\sim6\kpc$. The emission seen in the
north at $z\sim7.5\kpc$ is a point source (number 5 in Fig.~3).}
\label{fig:surf-b}
\end{figure*}

It is clear from Fig.~\ref{fig:wind-grey} that the diffuse emission is not
symmetric around the plane of the galaxy. The surface brightness
in a strip of width $0.10\deg$ parallel to the minor axis
(Fig.~\ref{fig:surf-b})
is initially higher to the south,
but then drops more rapidly than to the north. Beyond $2\kpc$
($\sim120\arcs$) from
the nucleus the northern wind is consistently brighter.
This asymmetry is also seen in the radio data of Seaquist \& Odegard
(1991). Within $100\arcs$ of the nucleus, the brightness profile at
$20\cm$ is brighter towards the south, while beyond $100\arcs$ the
north is brighter.

Emission can be traced out to $z\approx0.1\deg$ ($\equiv6\kpc$) from the
nucleus along the minor axis, before dropping into the noise.
The {\em Einstein} IPC estimate of emission extending to $\sim9\arcm$
from the nucleus was due to the inclusion of a point source (source 5
in Table~\ref{tab:pspc-srcs}) in the
diffuse emission (see Fig. 3 in Fabbiano 1988).

\subsection{Comparison with \hi distribution}
\begin{figure*}
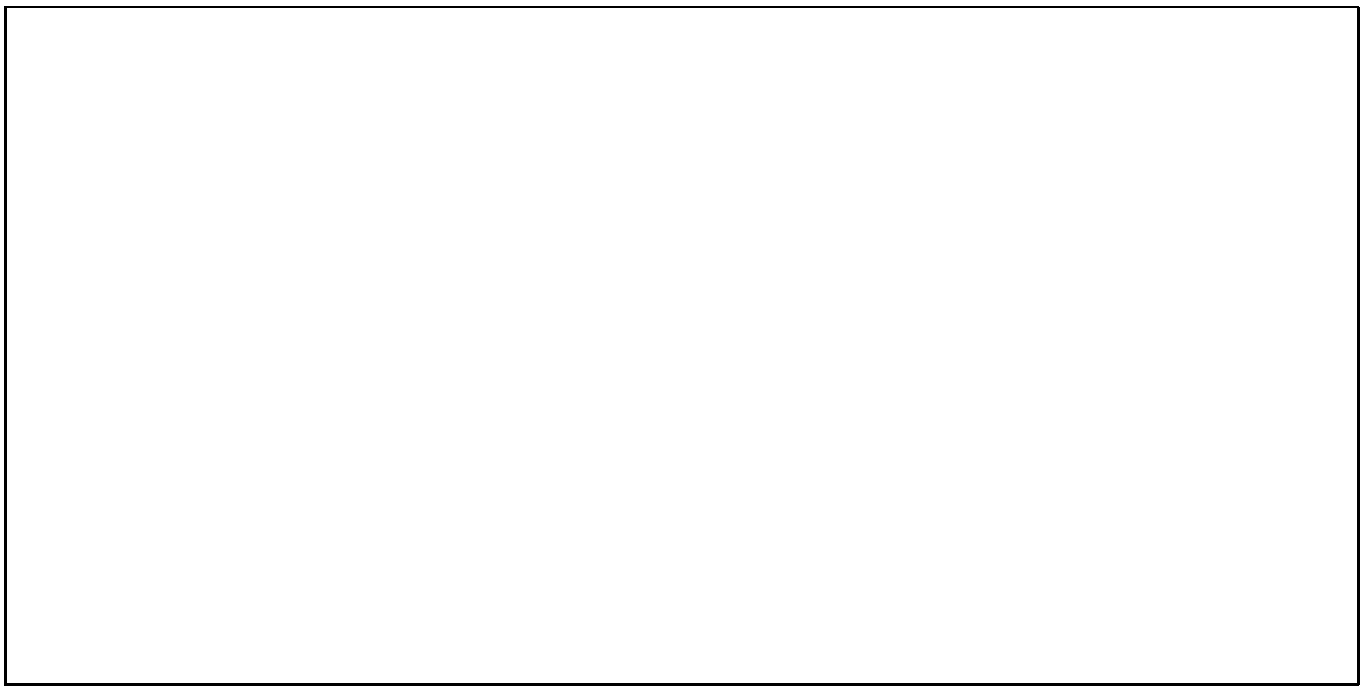

\picplace{9cm}
\caption{Comparison between X-ray and \hi distributions.
A greyscale \mbox{X-ray} image (lowest tone corresponding a flux
of $1.44\times10^{-3} {\rm counts} \ps$ arcmin$^{-2}$), overlaid with
contours of \hi column density (adapted from Yun \etal (1993)). The
contours correspond to $2.7\times10^{19} \pcm2$ times 1, 2, 3, 4, 6, 10,
15 and 25.}
\label{fig:h1-xray}
\end{figure*}

As can be seen in Fig.~\ref{fig:h1-xray}, the X-ray emission appears to
\mbox{anti-correlate} with the large scale distribution of \hi surrounding
M82. To the north-east, the \mbox{X-ray} emission appears to be bounded by the
northern tidal streamer. Yun \etal (1993) claim this to be M82's tidally
disrupted outer \hi disk. To the north-west, another streamer of \hi intrudes
onto the \mbox{X-ray} distribution on the eastern edge of regions n4 and n5.
This northern \hi has a velocity consistent with being on the
far side of M82, as is the northern wind.
The southern wind appears confined between the clump of hydrogen to the
south-east and the beginnings of the southern tidal streamer to the south-west.
The south-eastern \hi clump shows a broad blueshifted line wing which Yun \etal
(1993) note may be due to the impact of a wind.
The \hi in the tidal streamers could provide a natural obstacle for
the wind, constraining its expansion. The northern and southern
streamers each contain $\sim6\times10^{7} \Msol$, similar to the mass
of material we infer below for the soft X-ray emitting material
in the wind;
so the \hi could potentially form a significant barrier for the wind.

In the inner regions, Fig.~\ref{fig:h1-xray}, shows significant amounts
of \hi in the region occupied by the optical filamentation and inner wind.
The inner \hi displays a velocity gradient along the minor axis in
the same sense as the ${\rm H\alpha}$ emission, hence the \hi probably
consists of material swept out of the disk by the wind. As is discussed
below, the X-ray spectra show signs of excess absorption.

\label{sec:windres}
\subsection{Wind parameters}
The results of the spectral fitting to the wind regions are given in
Table~\ref{tab:wind-res}. Contamination of the spectral
properties of the wind by the nuclear point source has
been allowed for as discussed in section~\ref{sec:contam}.
As can be seen from Fig. 3
there is little wind emission beyond regions n8 and s6, and
no useful spectral parameters could be derived for these outermost
strips.

The absorbing column is found to decrease as the distance from the plane
of the galaxy increases. Only in the south does the
column drop to the Stark (1992) value of $4.0\pm{0.5}\times10^{20} \pcm2$.
It can be seen from
Fig.~\ref{fig:h1-xray} that, on the basis of the \hi distribution,
excess $\nH$ would be expected to extend only to $z\sim2\kpc$,
and the magnitude of the observed excess for the north ($\sim3\times10^{20} \pcm2$)
is larger than expected, except in the nuclear region.
Absorption in the ROSAT band arises predominantly from
He, C, N, and O rather than H{\sc i}, hence, if the absorbing gas has
a low metallicity such as is inferred for the hot \mbox{X-ray} emitting
gas, the absorbing masses required at large heights above the plane are several $10^{7}$
to several $10^{8} M_{\odot}$.

\begin{figure*}
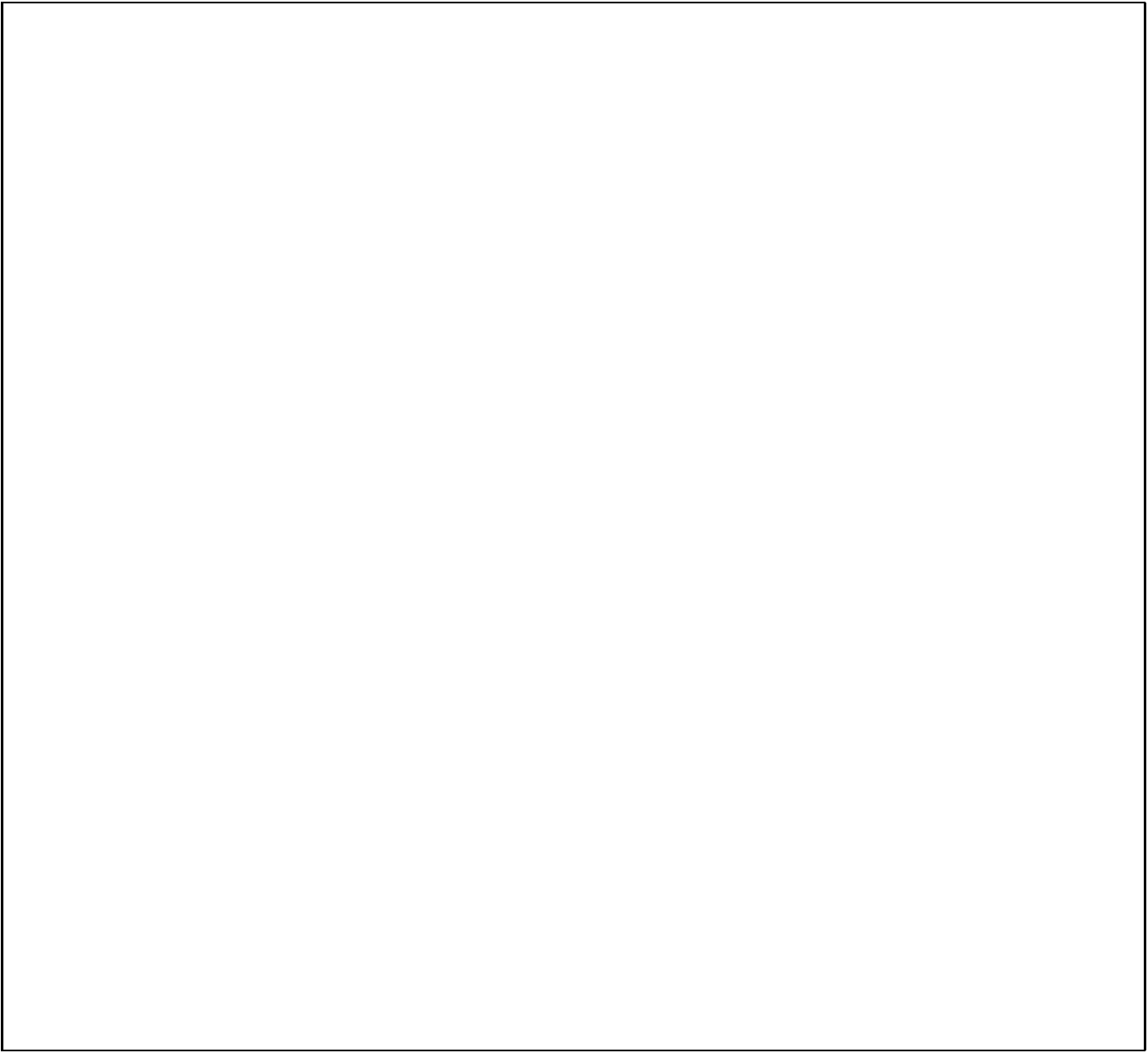

\picplace{16.5cm}
\caption{Error ellipses for column against temperature at 68\% confidence in two
interesting parameters for the wind regions.
Regions n7 and n8 are peculiar, as discussed in Sect.~4.5.
The dotted line shows the Stark column. For the northern regions (n1-n6) it is clear
that: a) the altering the temperature will not remove the need for excess absorption,
and b) the temperatures for the innermost regions are well determined.}
\label{fig:coltemp}
\end{figure*}

Fig.~\ref{fig:coltemp} shows 68\% confidence error ellipses for column and
temperature. Excess absorption is required to the north, although the column
for the south drops to close to the Stark value. The temperatures are well
constrained, and do not depend strongly on the fitted column. The origin
of the excess absorption to the north remains to be determined.

Temperature and density both decrease with increasing distance along
the minor axis $z$, although the temperature drop is small
(Figs ~\ref{fig:temp} -- ~\ref{fig:dens}). The density
is initially higher to the south, but then drops below
the density to the north beyond $\sim2\kpc$, as indicated by
the surface brightness profiles (Fig.~\ref{fig:surf-b}).

\begin{table*}
\caption[]{Spectral fits to the wind regions. The Stark column is
$0.40\pm{0.05}\times10^{21} \pcm2$. The metallicity is frozen at $0.05
Z_{\odot}$.}
\begin{flushleft}
\begin{tabular}{lllll}
\noalign{\smallskip}
\hline
\noalign{\smallskip}
Region & Counts & \multicolumn{3}{l}
	{Raymond-Smith parameters} \\
&& $\nH$ & T & EM \\
&&  (\tpow{21} $\pcm2$) & ($\keV$) & (\tpow{56} $\cm^{3}$ / $10 \kpc^{2}$) \\
\noalign{\smallskip}
\hline
\noalign{\smallskip}
n1 & $ 1714\pm{43}$  
	& $ 1.26^{+0.51}_{-0.21} $ & $ 0.65^{+0.04}_{-0.06} $ &
	 $ 44.7^{+10.2}_{-3.6} $ \\
n2 & $ 810\pm{30}$   
	& $ 0.85^{+0.13}_{-0.10} $ & $ 0.44^{+0.04}_{-0.04} $ &
	 $ 26.0^{+4.7}_{-3.5} $ \\
n3 & $ 579\pm{26}$   
	& $ 0.73^{+0.11}_{-0.09} $ & $ 0.41^{+0.04}_{-0.04} $ &
	 $ 20.0^{+3.9}_{-3.1} $ \\
n4 & $ 416\pm{23}$   
	& $ 0.73^{+0.15}_{-0.11} $ & $ 0.41^{+0.05}_{-0.05} $ &
	 $ 13.9^{+3.5}_{-2.6} $ \\
n5 & $ 295\pm{20}$   
	& $ 0.96^{+0.63}_{-0.21} $ & $ 0.32^{+0.05}_{-0.12} $ &
	 $ 16.3^{+10.4}_{-4.3} $ \\
n6 & $ 198\pm{17}$   
	& $ 0.94^{+0.57}_{-0.24} $ & $ 0.33^{+0.07}_{-0.07} $ &
	 $ 9.9^{+6.5}_{-3.2} $ \\
n7 & $ 116\pm{14}$   
	& $ 0.64^{+0.39}_{-0.18} $ & $ 0.76^{+0.20}_{-0.15} $ &
	 $ 2.4^{+0.4}_{-0.3} $ \\
n8 & $ 107\pm{14}$   
	& $ 0.61^{+0.69}_{-0.19} $ & $ 0.89^{+0.27}_{-0.14} $ &
	 $ 2.3^{+0.4}_{-0.3} $ \\
n9 & $ 40\pm{12}$   
	& $-$ & $-$ &
	 $-$ \\
\noalign{\smallskip}
\hline
\noalign{\smallskip}
s1 & $ 2859\pm{54}$   
	& $ 0.89^{+0.06}_{-0.05} $ & $ 0.60^{+0.02}_{-0.02} $ &
	 $ 69.6^{+3.3}_{-2.9} $ \\
s2 & $ 1494\pm{40}$   
	& $ 0.84^{+0.08}_{-0.07} $ & $ 0.56^{+0.03}_{-0.03} $ &
	 $ 38.5^{+2.9}_{-2.5} $ \\
s3 & $ 621\pm{27}$   
	& $ 0.61^{+0.08}_{-0.06} $ & $ 0.52^{+0.05}_{-0.05} $ &
	 $ 14.7^{+2.1}_{-1.5} $ \\
s4 & $ 246\pm{19}$   
	& $ 0.74^{+0.30}_{-0.16} $ & $ 0.44^{+0.09}_{-0.07} $ &
	 $ 7.6^{+3.1}_{-1.8} $ \\
s5 & $ 185\pm{17}$   
	& $ 0.50^{+0.16}_{-0.11} $ & $ 0.45^{+0.11}_{-0.08} $ &
	 $ 4.6^{+1.9}_{-1.1} $ \\
s6 & $ 141\pm{16}$   
	& $ 0.52^{+0.16}_{-0.13} $ & $ 0.33^{+0.09}_{-0.06} $ &
	 $ 5.3^{+2.4}_{-1.8} $ \\
s7 & $ 115\pm{15}$   
	& $-$ & $-$ &
	 $-$ \\
s8 & $ 86\pm{14}$   
	& $-$ & $-$ &
	 $-$ \\
s9 & $ 62\pm{13}$   
	& $-$ & $-$ &
	 $-$ \\
\noalign{\smallskip}
\hline
\end{tabular}
\end{flushleft}
\label{tab:wind-res}
\end{table*}

Under our assumed cylindrical geometry, it is possible to derive further
useful gas parameters (see Table~\ref{tab:wind-pars}). We assume a distance
of $3.63 \Mpc$ to M82 throughout. The volume $V$ is derived from the
geometry, of which the emitting gas is assumed to
occupy some fraction $\eta$ (the filling factor of the hot gas).
The fitted emission measure then
equals $\eta \, n^{2}_{e} \, V$, and
(assuming an ionised hydrogen plasma for simplicity)
the mean electron density, total gas mass
$M_{\mbox{\small gas}}=m_{\rm H} \,  n_{e} \, V$,
thermal energy $E_{\mbox{\small th}}=3 \, n_{e}\, V \,\k T$,
bulk kinetic energy $K_{\mbox{gas}}=\frac{1}{2} \,
M_{\mbox{\small gas}} v^{2}_{\rm gas}$, cooling timescale
$t_{\mbox{\small cool}}=E_{\mbox{\small th}} / L_{\rm X}$,
mass deposition rate
$\dot{M}_{\mbox{\small cool}}=M_{\mbox{\small gas}} /
t_{\mbox{\small cool}}$, and sound speed
$C_{\rm sound}=\sqrt(2 \, \gamma \, \k \, T \, / m_{\rm H})$,
can then be calculated.
The intrinsic (\ie corrected for
both galactic and intrinsic absorption) X-ray luminosity
$L_{X-in}$ in the {\em ROSAT} band is also given.

\begin{table*}
\caption[]{Derived gas parameters for the wind, assuming a distance of 
$3.63 \Mpc$ to M82. $\eta$ is the
volume filling factor of the gas. All parameters have been derived assuming $\eta=1$. Conversion 
factors to arbitrary $\eta$ are given. $v_{1000}$ is the outflow velocity of
the \mbox{X-ray} emitting gas in units of $1000 \kmps$, which may not be the same as the wind velocity. }
\begin{flushleft}
\begin{tabular}{llllllllll}
\noalign{\smallskip}
\hline
\noalign{\smallskip}
{\scriptsize Region} & {\scriptsize Volume} & {\scriptsize $n_{\rm e}$} &
	{\scriptsize $M_{\rm gas}$} &
	{\scriptsize $E_{\rm th}$} & {\scriptsize $C_{\rm sound}$} &
	{\scriptsize $K_{\rm gas}$} &
	{\scriptsize $L_{\rm X-in}$} &
	{\scriptsize $t_{\rm cool}$} & {\scriptsize $\dot{M}_{\rm cool}$} \\
& {\scriptsize(\tpow{65}$\cm^{3}$)} &
	{\scriptsize (\tpow{-3} $\pcc$)} & {\scriptsize($10^{6} M_{\odot}$)} &
	{\scriptsize($10^{55}$ ergs)} & {\scriptsize $\kmps$} &
	{\scriptsize($10^{55}$ ergs)} &
	{\scriptsize($10^{38} \ergps$)} & {\scriptsize(Myr)} &
	{\scriptsize($\Msol yr^{-1}$)} \\
&& {\scriptsize($\times1/\sqrt{\eta}$)} & {\scriptsize($\times\sqrt{\eta}$)} &
	{\scriptsize($\times\sqrt{\eta}$)} &
	& {\scriptsize($v_{1000}^{2}\times\sqrt{\eta}$)} &
	{\scriptsize($0.1-2.4 \keV$)} &
	{\scriptsize($\times\sqrt{\eta}$)} & \\
\noalign{\smallskip}
\hline
\noalign{\smallskip}
n1 & 5.85
	& $ 31.7^{+3.6}_{-1.4} $ &
	15.4 & 5.6 & 460
	& 15.4 & 33.8 
	& 520 & 0.030 \\
n2 & 5.85
	& $ 24.2^{+2.2}_{-1.6} $ &
	11.8 & 2.9 & 370
	& 11.8 & 16.2 
	& 560 & 0.021 \\
n3 & 5.85
	& $ 21.2^{+2.1}_{-1.7} $ &
	10.4 & 2.4 & 360
	& 10.4 & 11.8 
	& 640 & 0.016 \\
n4 & 5.85
	& $ 17.7^{+2.2}_{-1.6} $ &
	8.6 & 2.0 & 360
	& 8.6 & 8.3 
	& 750 & 0.011 \\
n5 & 5.85
	& $ 19.2^{+6.1}_{-2.5} $ &
	9.4 & 1.7 & 320
	& 9.4 & 8.3 
	& 640 & 0.015 \\
n6 & 4.17
	& $ 17.7^{+5.8}_{-2.9} $ &
	6.2 & 1.1 & 330
	& 6.2 & 5.2 
	& 690 & 0.009 \\
n7 & 4.15
	& $ 8.7^{+0.7}_{-0.5} $ &
	3.0 & 1.3 & 490
	& 3.0 & 1.9 
	& 2160 & 0.001 \\
n8 & 5.85
	& $ 7.1^{+0.6}_{-0.4} $ &
	3.5 & 1.7 & 540
	& 3.5 & 1.8 
	& 3050 & 0.001 \\
\noalign{\smallskip}
\hline
\noalign{\smallskip}
\multicolumn{3}{l}{{\scriptsize Sub-total (North)}}
	&  68.4 & 18.6 &
	& 68.4 &  87.2  &  & 0.104\\
\noalign{\smallskip}
\hline
\noalign{\smallskip}
s1 & 5.85
	& $ 39.6^{+0.9}_{-0.8} $ &
	19.3 & 6.4 & 440
	& 19.3 & 51.0 
	& 400 & 0.048 \\
s2 & 5.85
	& $ 29.4^{+1.1}_{-0.9} $ &
	14.4 & 4.5 & 430
	& 14.4 & 27.5 
	& 520 & 0.028 \\
s3 & 5.85
	& $ 18.2^{+1.3}_{-0.9} $ &
	8.9 & 2.6 & 410
	& 8.9 & 10.1 
	& 800 & 0.011 \\
s4 & 5.85
	& $ 13.0^{+2.7}_{-1.6} $ &
	6.4 & 1.6 & 370
	& 6.4 & 4.7 
	& 1050 & 0.006 \\
s5 & 5.85
	& $ 10.1^{+2.1}_{-1.3} $ &
	5.0 & 1.2 & 380
	& 5.0 & 2.9 
	& 1370 & 0.004 \\
s6 & 4.55
	& $ 10.9^{+2.5}_{-1.9} $ &
	4.1 & 0.8 & 330
	& 4.1 & 2.7 
	& 880 & 0.005 \\
\noalign{\smallskip}
\hline
\noalign{\smallskip}
\multicolumn{3}{l}{{\scriptsize Sub-total (South)}} 
	& 58.1 & 17.0 &  & 58.1 & 98.9  
	& & 0.102 \\
\noalign{\smallskip}
\hline
\noalign{\smallskip}
\multicolumn{3}{l}{{\scriptsize Total}}
	& 126.4 & 35.6 &  & 126.4 & 186.2 
	&  &  0.206 \\
\noalign{\smallskip}
\hline
\end{tabular}
\end{flushleft}
\label{tab:wind-pars}
\end{table*}

In order to quantify the trends in the data, particularly in
the behaviour of the temperature and density with increasing
distance along the wind, we perform weighted \mbox{least-squares}
fits to the data for north and south
separately. We also regress temperature against density using weighted
orthogonal regression
(Feigelson \& Babu 1992), allowing for the significant
errors on both axes, using the package ODRPACK (Boggs \etal 1992).

Table~\ref{tab:reg-fits} gives the fitted slopes, while
the data and fitted lines are plotted in
Figs. ~\ref{fig:temp} - \ref{fig:temp-dens}.
The fits used data for regions n1-n6 and s1-s6 only;
regions n7 \& n8 were excluded as they clearly deviate
from the general trend in the North.

The elevated temperatures of n7 and n8 are difficult to explain.
We have checked that the two point sources which fall in the vicinity
have been effectively excluded from the data. One obvious possibility
is that the temperature rise is due to a shock, however, in this case
the density would also be
expected to rise, whereas it appears {\it lower} than expected
from the trend of the inner six northern regions.
A hardness map shows a lack of soft flux at the edges of the wind in
regions n7 and n8, with no corresponding lack of hard flux,
but the regions of reduced soft flux do not seem to correspond to
areas of excess \hi and hence higher absorption.
We have investigated the possibility of  the excess hard flux being
due to the energy dependent scattering from inner regions of
the wind, but such contamination from one region into the next
is at too low a level, {\em decreases} in importance with $z$,
and is not strongly energy dependent. Also, it should be noted
that there is no corresponding temperature rise in the south.

\begin{table*}
\caption[]{Results of the linear regression applied separately to the data
from both north and south winds as described in the text. $z$ is the 
distance along the minor axis. Results are given for both the contamination ``corrected''
and uncorrected data to demonstrate the effect the contamination has.
To assess the effect of the chosen geometry, results for a truncated
conical geometry with radius on the major axis $r_{\rm base}=0.03\deg$
and an opening angle of $50\deg$ are also shown.}
\begin{flushleft}
\begin{tabular}{llllll}
\noalign{\smallskip}
\hline
\noalign{\smallskip}
\multicolumn{6}{l}{Cylinder, radius $r=0.05\deg$ Corrected
	for nuclear contamination of the wind.} \\
\noalign{\smallskip}
\hline
\noalign{\smallskip}
Relationship & Wind & \multicolumn{2}{l}{Slope} 
	& \multicolumn{2}{l}{Intercept} \\
 & & & {\scriptsize 95\% confidence interval} & 
	& {\scriptsize 95\% confidence interval} \\
\noalign{\smallskip}
\hline
\noalign{\smallskip}
T:$n_{e}$  & North & $1.047\pm{0.206}$  & 0.474 to 1.620 
	& $1.374\pm{0.335}$ & 0.445 to 2.303 \\
  & South & $0.262\pm{0.058}$ & 0.100 to 0.423 
	& $0.146\pm{0.087}$ & $-0.097$ to 0.389 \\
\noalign{\smallskip}
T:$z$  & North & $-0.474\pm{0.078}$ & $-0.691$ to $-0.258$ 
	& $1.208\pm{0.251}$ & 0.511 to 1.906 \\
  & South & $-0.227\pm{0.060}$ & $-0.394$ to $-0.061$ 
	& $0.458\pm{0.186}$ & 0.057 to 0.974 \\
\noalign{\smallskip}
$n_{e}$:$z$  & North & $-0.467\pm{0.045}$ & $-0.590$ to $-0.343$ 
	& $0.117\pm{0.144}$ & $-0.517$ to 0.283 \\
  & South & $-0.836\pm{0.108}$ & $-1.137$ to $-0.535$ 
	& $1.095\pm{0.334}$ & 0.168 to 2.022 \\
\noalign{\smallskip}
\hline
\noalign{\smallskip}
\multicolumn{6}{l}{Cylinder, radius $r=0.05\deg$. No
	contamination correction.} \\
\noalign{\smallskip}
\hline
\noalign{\smallskip}
Relationship & Wind & \multicolumn{2}{l}{Slope} 
        & \multicolumn{2}{l}{Intercept} \\
 & & & {\scriptsize 95\% confidence interval} & 
        & {\scriptsize 95\% confidence interval} \\
\noalign{\smallskip}
\hline
\noalign{\smallskip}
T:$n_{e}$  & North & $0.821\pm{0.119}$ & 0.490 to 1.151 
	& $1.074\pm{0.195}$ & 0.532 to 1.616 \\
  & South & $0.185\pm{0.052}$ & 0.041 to 0.329 
	& $0.079\pm{0.078}$ & $-0.137$ to 0.295 \\
\noalign{\smallskip}
T:$z$  & North & $-0.467\pm{0.058}$ & $-0.628$ to $-0.305$ 
	& $1.236\pm{0.188}$ & 0.715 to 1.758 \\
  & South & $-0.188\pm{0.039}$ & $-0.295$ to $-0.081$ 
	& $0.384\pm{0.120}$ & $-0.052$ to 0.715 \\
\noalign{\smallskip}
$n_{e}$:$z$  & North & $-0.568\pm{0.072}$ & $-0.769$ to $-0.368$ 
	& $-0.189\pm{0.233}$ & $-0.457$ to 0.835 \\
  & South & $-0.922\pm{0.114}$ & $-1.239$ to $-0.606$ 
	& $1.353\pm{0.351}$ & 0.378 to 2.328 \\
\noalign{\smallskip}
\hline
\noalign{\smallskip}
\multicolumn{6}{l}{Truncated cone, radius $r_{\rm base}=0.03\deg$,
	opening angle $\theta_{\rm op}=50\deg$, contamination corrected.} \\
\noalign{\smallskip}
\hline
\noalign{\smallskip}
Relationship & Wind & \multicolumn{2}{l}{Slope} 
        & \multicolumn{2}{l}{Intercept} \\
 & & & {\scriptsize 95\% confidence interval} & 
        & {\scriptsize 95\% confidence interval} \\
\noalign{\smallskip}
\hline
\noalign{\smallskip}
T:$n_{e}$  & North & $0.641\pm{0.135}$ & 0.266 to 1.0170 
	& $0.695\pm{0.207}$ & 0.121 to 1.270 \\
  & South & $0.204\pm{0.051}$ & 0.063 to 0.346 
	& $0.037\pm{0.072}$ & $-0.164$ to 0.237 \\
\noalign{\smallskip}
T:$z$  & North & $-0.476\pm{0.089}$ & $-0.724$ to $-0.227$ 
	& $1.223\pm{0.284}$ & 0.436 to 2.010 \\
  & South & $-0.228\pm{0.068}$ & $-0.418$ to $-0.038$ 
	& $0.459\pm{0.213}$ & $-0.132$ to 1.049 \\
\noalign{\smallskip}
$n_{e}$:$z$  & North & $-0.762\pm{0.027}$ & $-0.837$ to $-0.687$ 
	& $0.887\pm{0.086}$ & 0.649 to 1.125 \\
  & South & $-1.103\pm{0.118}$ & $-1.431$ to $-0.775$ 
	& $2.023\pm{0.365}$ & 1.010 to 3.036 \\
\noalign{\smallskip}
\hline
\end{tabular}
\end{flushleft}
\label{tab:reg-fits}
\end{table*}

\begin{figure*}
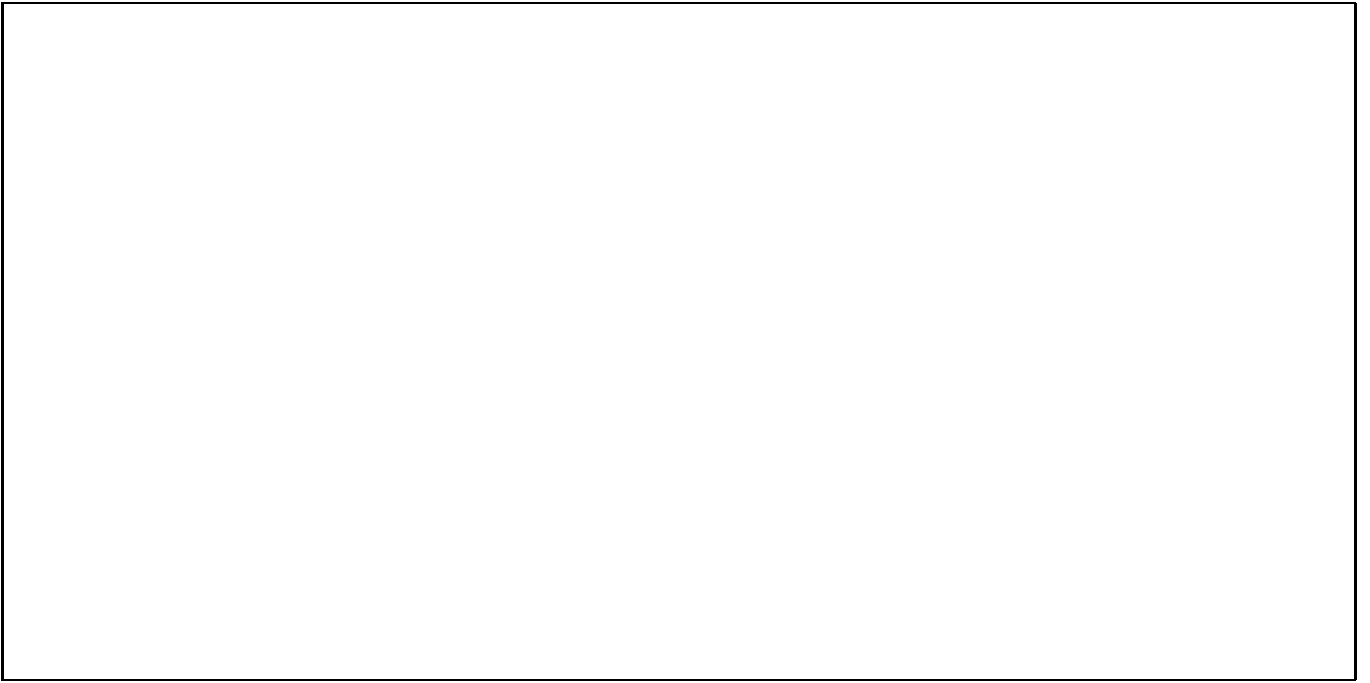

\picplace{9cm}
\caption[]{Absorbing column against minor axis distance for
Northern (crosses) and Southern (diamonds) winds. The column is in
units of $10^{21} \pcm2$. The Stark (1992) column is $0.40\pm{0.05}$ in these
units. As the northern side of M82 is inclined away from us, the
initially higher column to the north is entirely natural.}
\label{fig:hcol}
\end{figure*}

\begin{figure*}
\picplace{9cm}
\caption[]{Temperature against minor axis distance for Northern
(data: crosses, regression line: dashed)
and Southern(data: diamonds, regression line: solid) winds.
Only the first six points (n1 -- n6 and s1 -- s6)
are used in the regressions.}
\label{fig:temp}
\end{figure*}

\begin{figure*}
\picplace{9cm}
\caption[]{Derived density against minor axis distance for Northern
(data: crosses, regression line: dashed)
and Southern (data: diamonds, regression line: solid) winds.
Only the first six points (n1 -- n6 and s1 -- s6) are used in the
regressions.}
\label{fig:dens}
\end{figure*}

\begin{figure*}
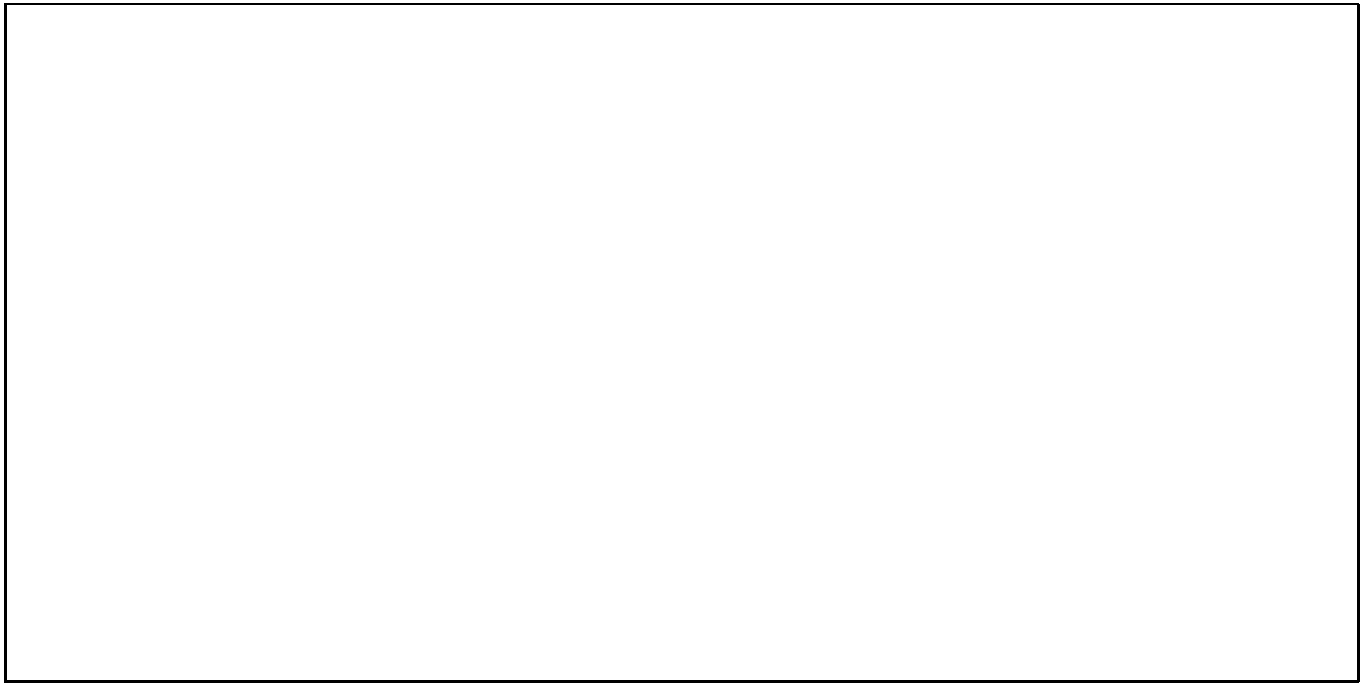

\picplace{9cm}
\caption[]{Temperature against density for Northern (data: crosses,
regression line: dashed)
and Southern (data: diamonds, regression line: solid) winds.
For an adiabatically expanding gas the slope of the $\log T : \log \rho$ regression
line would be $\gamma-1 = \frac{2}{3}$.}
\label{fig:temp-dens}
\end{figure*}

A systematic error in the background subtraction could
possibly mimic a real trend with increasing distance along the wind due to
the increasing importance of the background as the surface brightness of
the emission decreases. To check the effects of this, the
analysis was repeated for
backgrounds 5\%  over and undersubtracted with respect to the ideal
background described above. The fitted parameters were within $1\sigma$
of those from the standard background in all cases. Hence over or
undersubtraction is not a serious problem.

\subsection{The effect of the assumed geometry}
\label{sec:assum_geom}
As discussed in Sect.~\ref{sec:an_wind},
M82's \mbox{X-ray} emission is not obviously well approximated by either a
conical or a cylindrical outflow. In addition, the
asymmetry between north and south makes the choice of a
consistent geometry difficult. The inclination of the galactic
plane to our line of sight will also blur any results,
by superposition of physically differing regions, even if the plasma's
properties do vary only with $z$. It is not unreasonable to
expect variation perpendicular to the minor axis, leading
to further superposition of different components along the line
of sight. We checked for this by performing spectral fits for
regions n3, n4, s3 \& s4,
binning the emission into eastern, central and western spectra.
The resulting temperatures across the wind fell
within $1\sigma$ of each other, indicating that cross-wind variations
are not a major effect.

Let us suppose that the plasma properties vary not with
$z$, but with radius from the galactic centre, as in a spherical or conical
outflow. Our
derived spectral properties, using a cylindrical geometry,
will then differ from the true properties. For a
conical distribution, the degree to which the fitted parameters
deviate from the true parameters depends on the opening angle of the
cone. The fitted parameters will be some flux-weighted
average of the various components of different $r$ that fall within
a slice at constant $z$. The effect will be worst for the inner regions,
and for large cone opening angle,
but will be small at large
$z$. For a parameter $F$ that decreases with $r$,
the fitted parameter $F$ at some $z$ will always be lower than
the true value for $r=z$, due to the incorporation of flux from
regions of greater $r$ and hence lower $F$. This will have the
effect of flattening the slope of any real trend in the data as the
discrepancy between true and fitted values is less at large $r$.

In the present case, given the
flatness of the temperature profile,
the true temperature will not vary much from
the fitted values. The emission measure will not be too far out
either, given that the higher surface brightness emission along
the minor axis dominates the
fitted emission measure (\ie the inclusion of the lower surface
brightness emission further from the minor axis in our cylindrical geometry
has rather little effect).
The major source of bias is the volume, which we would overestimate
at small $r$, and underestimate at large $r$.
However, it should be borne in mind that the derived gas density depends
only on the inverse square root of the assumed volume,
$n_{\rm e} \propto \sqrt({\rm EM} / V )$.

In conclusion, if the soft X-ray emission does have a rather more divergent
geometry than we have assumed, then the true temperatures will be similar to
those obtained, while the density will drop off faster than our result, the
inner densities being higher and the outer densities lower than those we have
derived.

In order to test the magnitude of these effects, the standard analysis above
was repeated treating the emission as arising from  two inverted truncated cones
of radius $r_{0}=0.03 \deg$ on the galactic major axis and \mbox{semi-opening}
angle $\theta = 25 \deg$, chosen by eye to give an reasonable approximation to
the observed emission in the North (see Fig.~\ref{fig:cone}). The results of
this on the derived trends in $n$ and $T$, can be seen in
Table~\ref{tab:reg-fits}. The slope of the $T:z$ relation is essentially
unchanged, whilst the $n_{\rm e}:z$ trend becomes steeper, as expected.
The implications of this will be discussed below.

\section{Comparison with wind models}
\label{sec:discuss}
Given previous claims that the X-ray emission from M82 is consistent
with that expected from the
adiabatic expansion of a free flowing wind (Fabbiano 1988;
Bregman \etal 1995), in particular that the density derived from
the surface brightness falls off as $\sim r^{-2}$, it is worth
investigating whether our spectral results are consistent with this idea.

\subsection{Chevalier \& Clegg's adiabatic wind}
The simplest useful model of a galactic wind is
that of Chevalier \& Clegg (1985). This is just a spherically-symmetric
outflow from a region of constant mass and energy injection (the
starburst) ignoring the effects of gravity, radiative cooling and
the presence of any ambient medium. The hot gas smoothly
passes through a sonic transition at the radius of the starburst
region, and then becomes a supersonic outflow which cools
adiabatically. Provided the kinetic energy supplied by the
numerous SN and stellar winds within the starburst is efficiently
thermalised, the temperature of the hot gas within the starburst region will
be $\sim 10^{8} \K$ for reasonable mass and energy injection
rates, making the neglect of the effects of gravity and radiative
cooling valid. For regions well outside the starburst region,
$r \gg R_{\star}$, the wind density $\rho \propto r^{-2}$,
wind temperature $T \propto r^{-4/3}$ and thermal pressure
$P_{\rm th} \propto r^{-10/3}$. McCarthy \etal (1987) compared
variation of pressure with radius in the optical filamentation
with the Chevalier \& Clegg (CC) model, under the assumption that
the thermal pressure in the filaments would equal the total
(thermal plus ram) pressure in the wind, achieving a good match within
a kiloparsec of the nucleus.
Fabbiano (1988) reported in a reanalysis of the {\em Einstein}
IPC data, that the radial distribution of the X-ray emission was
consistent with $\rho \propto r^{-2}$, \ie a free-flowing wind.

The Chevalier \& Clegg model assumes a spherical outflow, in contrast
to the cylindrical geometry we have adopted.
What would we expect to see from a free wind in a more
generalised outflow geometry, \eg a bubble that has broken out of the
disk of the galaxy and now allows free escape of the wind material?
Assuming a constant mass loss rate, $\dot{M}
= \rho_{r} A_{r} v_{r}$, where $A_{r}$ and $v_{r}$ are
the \mbox{cross-sectional} area and the velocity of the flow, and
$A_{r}$ of the form $A_{r} \propto r^{\beta}$, together with a constant outflow
velocity, it follows that $\rho_{r} \propto r^{-\beta}$. For a cylinder,
$\beta=0$, hence $\rho$ is constant. For a sphere or a cone of
constant opening angle, $\beta=2$. The density for a cone is just
a constant ratio higher
($4\pi/ \Omega$, where $\Omega$ is the opening angle in
steradians)
than that for a spherical wind for a constant $\dot{M}$.
Obviously we can produce any $\rho_{r}$ and retain the concept
of the emission as arising from a free wind, by choosing the
appropriate geometry. However, for isentropic gas, the temperature
$T_{r} \propto \rho^{\gamma-1}_{r}$, where $\gamma=5/3$ in the present case.
In the absence of cooling (a good approximation given
that the outflow timescale is $\sim10^{6}-10^{7} \yr$ while the
cooling timescales are $\sim10^{9} \yr$) a free wind
would expand adiabatically. Hence we expect a
$\log T : \log \rho$ graph to have a slope of
$\gamma-1=\frac{2}{3}$ for {\em any} free wind irrespective of
geometry. Inspection of Table~\ref{tab:reg-fits} shows that
although the northern emission is consistent with this,
the southern emission is inconsistent
at greater than the 95\% confidence level, in the sense that
the temperature drops too slowly relative to the density
-- \ie the entropy {\it rises} outwards.

Note that if the absorbing columns in the northern outer regions
are overestimated, then the real temperatures for these regions will
be higher, reducing the temperature gradient. This would make the north
less isentropic and reduce the difference between north and south.
However the fits clearly require a higher column than Stark for these
regions.

As discussed in Sect.~\ref{sec:assum_geom}, our use of a cylindrical
geometry will lead us to underestimate the slope of the density
profile if the emission comes from a conical outflow, whilst our temperature
estimate is quite robust.
This means that the inconsistency of the southern wind with
an adiabatic outflow can only be {\em accentuated} if the flow
diverges. The slopes for the re-analysis with a
truncated conical geometry confirm this, the
southern emission being less isentropic than for the cylindrical analysis.


Bearing in mind the geometry issue, we can attempt a more
quantitative comparison between the observed temperatures and
densities and the CC model, using the forms
for $\rho$ and $T$ from CC, and the scaling relationships for the
mass and energy injection given by Heckman \etal (1993).
These use the predicted deposition of
mass, kinetic energy and momentum from a starburst calculated by
Leitherer \etal (1992). For a constant star formation rate over
a period of $5\times10^{7} \yr$, solar metallicity and a normal
Salpeter IMF extending up to $100 \Msol$, the various
injection rates are related to the starburst bolometric luminosity
by:
\begin{eqnarray}
\dot{E} & =  & 8\times10^{42} \, L_{\rm bol,11} \ergps \\
\dot{M} & =  & 3 \, L_{\rm bol,11} \Msol \pyr
\end{eqnarray}

\noindent where $L_{\rm bol,11}$ is the bolometric luminosity in units of
$10^{11} \Lsol$. For most starbursts $L_{\rm FIR}$, is the dominant
contributor to $L_{\rm bol}$ so it is a
reasonable approximation to equate $L_{\rm bol,11}$ with $L_{\rm FIR}$.
How valid are these scaling relations for M82?
Visual inspection of
Leitherer \etal's figures show that for all but the lowest
metallicities the injection rates are approximately
constant after $\sim5\times10^{6} \yr$, similar to the
age of M82's starburst (Rieke \etal, 1993).

Applying the above scaling relations to Chevalier \& Clegg's model
we obtain, for radii large compared to the starburst radius $R_{\star}$:
\begin{eqnarray}
n_{\rm e} (\pcc) & = & 8.14\times10^{-2}L_{\rm bol,11}
\left(\frac{r}{R_{\star}}\right)^{-2} 
\left(\frac{4\pi}{\Omega}\right)  \\
T (\keV) & = & 4.13 \left(\frac{r}{R_{\star}}\right)^{-4/3}
\left(\frac{4\pi}{\Omega}\right)^{2/3}
\end{eqnarray}

\noindent where $n_{\rm e}$ is the electron number density number
density, $T$ the temperature and $\Omega$ the total solid angle
through which the wind flows out.

Note that the temperature is independent of $L_{\rm bol}$, being a ratio of
the energy and mass injection rates. For a bolometric luminosity of
$4\times10^{10} \Lsol$ (Rieke \etal 1993) and a  characteristic radius,
$R_{\star}$, for the starburst of $200 \pc$, we can predict $\rho$ and $T$
at radii corresponding to the regions in
Fig.~\ref{fig:wind-reg} -- see Table~\ref{tab:models}.
Even for the innermost regions (n1 and s1) the CC model underestimates
the density by an order of magnitude.
Although the predicted temperature is almost equal to that observed near
the galactic centre, the
adiabatically expanding wind cools too quickly to match the PSPC data,
dropping to $\sim0.05 \keV$ in the outermost regions, whereas the
observed temperature along the minor axis is almost constant
at $0.4-0.5 \keV$ and even rises in regions $n7$ and $n8$ to
$\sim0.8 \keV$.

In summary, 
the densities and temperatures derived under our assumed cylindrical
geometry differ greatly from those predicted under a spherical geometry
by the CC model. Given an arbitrary outflow geometry, it is in principle
possible for the adiabatic wind model to reproduce the observed shallower
trend in density, however the departure of the southern wind from
constant entropy is a robust result which is incompatible
with the basic assumptions of the CC model.
We conclude, therefore, that the observed \mbox{X-ray}
emission cannot arise from a single phase, expanding wind.
The fact that the entropy actually {\it rises} with $z$ in the south,
means that the CC model cannot be saved by supposing that additional
material is entrained into the flow as it proceeds. Although this might
raise the density, it would cause the entropy to decline with $z$,
accentuating the disagreement with our results.

\subsection{A hydrostatic halo}
It has been suggested (W.~Pietsch, private communication) that the
extended \mbox{X-ray} emission around NGC 253 may be due not to a
wind or shocked cloud emission, but to a static halo. For M82
the optical emission line velocity data and
spectral index variations in the radio halo (McKeith \etal 1995;
Seaquist \& Odegard 1991) demonstrate conclusively the
presence of a galactic wind. Since the synchrotron
emission is very similar to the \mbox{X-ray} in extent, so
it is difficult to argue that the \mbox{X-ray} emission is
not associated with a wind.

Can we rule out a hydrostatic halo on the basis of the observed
\mbox{X-ray} properties? From the observed density
and temperature the mass within some radius $r$ for a
hydrostatic halo, assuming spherical symmetry, is:
\begin{equation}
M(<r) \approx 4.45\times10^{10} \, (\beta+\epsilon) \,
T_{\keV} \, r_{\rm kpc}  \Msol,
\end{equation}

\noindent where the observed temperature $T\propto r^{-\epsilon}$ and
density $n_{\rm e} \propto r^{-\beta}$. For regions n3 and s3 the
predicted mass within $2.2\kpc$ are $4\times10^{10} \Msol$ and
$5\times10^{10} \Msol$ respectively. From G\"{o}tz \etal (1990) the mass
within this radius, based on velocities measured in \hi, is
$\sim3\times10^{9} \Msol$, an order of magnitude lower.

Hence we can rule out any possibility that the X-ray emitting  gas is bound to
M82 as a static halo.

\subsection{Shocked clouds in a wind}
An alternative to emission from a free wind is emission
from shocked material embedded in such a wind.
Any clouds of denser material overrun by the wind, be they
fragmented remnants of the dense shell swept up by the wind
in its ``snow-plough'' phase or clouds in the ISM, will
be shock-heated. The ${\rm H\alpha}$ filamentation
seen along the minor axis has line ratios indicative of
material shocked to $\sim10^{4} \K$. Less dense material
would be heated to even higher temperatures, and could be the
source of the soft \mbox{X-rays} seen, rather than the emission
being due to the wind itself.

The temperature to which these clouds will be heated depends
on the speed of the shock driven into them by the wind. This
depends on the relative densities of the cloud and the wind,
and the wind velocity. In the case of strong shocks (Mach number
M $\gg$ 1) we
can ignore the thermal pressure of both the wind and the cloud,
and equate momentum flux across the shock. Given the mass and energy
injection rates, one expects a wind velocity $v_{\rm w} \approx 3000
\kmps$, whereas the sound speed in the \mbox{X-ray} emitting gas is $\sim400
\kmps$.  Clouds will be accelerated by the wind to varying extents
depending on their column density and individual histories, but
only to velocities of order hundreds of $\kmps$, as seen in the ${\rm H\alpha}$
filaments. Hence the strong shock approximation is not unreasonable.
The shock driven into the cloud will then have a velocity
$v_{\rm c} \approx v_{\rm w} \sqrt{\rho_{\rm w}/\rho_{\rm c}}$. The eventual
temperature of the cloud will be proportional to
$v^{2}_{\rm c}$.

\begin{table}
\caption[]{Predicted parameters of the \mbox{X-ray} emitting gas using the simple models
discussed in Section 5. The shocked cloud temperatures are calculated using
the derived electron densities from the northern wind
for unit filling factor. Predicted cloud
temperatures for the south are similar. Values assume spherical outflow for the wind.
For a conical outflow of total solid angle $\Omega$, the shocked cloud temperature
$T_{\rm c}$ and the CC wind density are $\propto 4\pi / \Omega$, and the CC
temperature $\propto (4\pi / \Omega)^{2/3}$.}
\begin{flushleft}
\begin{tabular}{lllll}
\hline\noalign{\smallskip}
Region & $z$ & \multicolumn{2}{l}{Chevalier \& Clegg} &
	Shocked clouds  \\
&& $n_{\rm e}$ & $T_{\rm w}$ & $T_{\rm c}$  \\
& (pc) & ($10^{-4}\pcc$)  & ($\keV$)& ($\keV$) \\
\noalign{\smallskip}
\hline\noalign{\smallskip}
1 & 945 & 14.6 & 0.52 & 1.58 \\
2 & 1575 & 5.3 & 0.26 & 0.74 \\
3 & 2205 & 2.7 & 0.17 & 0.43 \\
4 & 2835 & 1.6 & 0.12 & 0.31 \\
5 & 3465 & 1.1 & 0.09 & 0.19 \\
6 & 4095 & 0.8 & 0.07 & 0.15 \\
7 & 4725 & 0.6 & 0.06 & 0.23 \\
8 & 5355 & 0.5 & 0.05 & 0.22 \\
\noalign{\smallskip}
\hline
\end{tabular}
\end{flushleft}
\label{tab:models}
\end{table}

For a constant velocity, spherical or conical wind flowing into solid angle
$\Omega$, with constant mass injection rate $\dot{M}$, the wind density
$\rho_{\rm w} \propto r^{-2}$, and the shocked cloud temperature
\begin{equation}
T_{\rm c} =  6.15\times10^{-3} v_{1000} \,  \dot{M} \, r^{-2}_{\rm kpc}
  \, n^{-1}_{\rm c}\left(\frac{4\pi}{\Omega}\right) \, \keV,
\end{equation}

\noindent where $v_{1000}$ is the wind velocity in units of $1000 \kmps$,
and $\dot{M}$ in units of $M\sol \pyr$. We assume the postshock cloud
density is four times the preshock density, and that ionization,
dissociation, magnetic fields and radiative losses are negligible. Relaxing
the previous four assumptions would result in lower shocked cloud
temperatures.

Given $n_{\rm c} \approx 2n_{\rm e}$, the number densities derived
above, a wind velocity of $3000\kmps$ and $\dot{M}$ from Eq.~(2) we can predict
the temperature we expect assuming unit filling
factor and $\Omega=4\pi$ (Table~\ref{tab:models}).
From Fig.~\ref{fig:wind-grey} one can estimate the solid angle the wind flows
into as $\sim\pi$ steradians, raising $T_{\rm c}$ by a factor $\sim4$.
In the context of clouds in a wind, the filling factor should be substantially
less than unity. The ${\rm H\alpha}$ filaments have $\eta \sim 10^{-2}$
(McCarthy \etal 1987) in the inner kiloparsec, so it would
not be unreasonable to expect cloud filling factors of order
$10^{-2}-10^{-1}$ at larger distances from the nucleus. This would reduce
$T_{\rm c}$ by a factor $\sim3$--10. The net effect is that the predicted
temperatures are rather lower than those observed, but considering the
large uncertainties involved, this simple model must be regarded
as giving results consistent with observation.

Assuming that $\Omega$ and the filling factor do not vary with distance
along the wind,
the predicted temperatures (Table~\ref{tab:models})
drop off too quickly to match the observed
trend of $T$ with $z$. The steepness of the predicted temperature profile
could be due to the assumed geometry. As discussed in
Sect.~\ref{sec:assum_geom} the inner densities may be higher than
calculated. Higher inner cloud densities would lead to lower postshock
cloud temperatures and hence flatten the trend.

So, in summary, predicted postshock temperatures for a simple model where
the wind shock heats clouds are consistent with, if slightly lower than,
those observed, given the observed emission measure (density).

\subsection{Numerical models}
\label{sec:nummod}
Tomisaka \& Ikeuchi (1988) were the first to explicitly model the wind
in M82 using 2D hydrodynamical simulations. They found a roughly
cylindrical bipolar wind formed naturally for a constant mass and
energy input rate in the nucleus of 0.1 SN $\pyr$. However, they modelled
the ISM as a cold rotationally supported disk in which the angular velocity
was independent of the distance from the plane of the disk. This physically
unrealistic configuration creates a strong funnel along the
$z$-\mbox{axis} which strongly collimates the wind.
Tomisaka \& Bregman (1993) allowed the rotational velocity to decrease
exponentially away from the cold disk, into a hot low density halo.
This distribution still provides a strong cylindrical funnel for the
expansion of the wind at low $z$. Suchkov \etal (1994, hereafter SBHL)
provide a more realistic ISM
for their modelling of M82: a two component cold rotating dense disk and
non-rotating hot tenuous halo, and a starburst history incorporating the
milder mass and energy input from stellar winds before the more energetic
supernovae dominated stages. They find bipolar outflows form easily over
a wide range in different halo and disk conditions. Shocked halo gas at
temperatures $0.2-0.4\keV$ provides the majority of the \mbox{X-ray}
emission in their ``soft'' band ($0.1-2.2\keV$).

The eventual wind geometry depends on the initial gas distribution and
the mass and energy input history. For those models with ``mild'' early
winds (models A1 and A2 in SBHL) biconical outflows of opening
angle $\theta \sim 60-90\deg$ with dense disk material entrained along
the surface do occur naturally. Initially the mild wind creates a cavity
in the disk to the halo for the wind to escape, without substantially
damaging the disk. The wind then propagates outward in the halo, sweeping
it up and shocking it. In the later SN-driven stage of the starburst, the
more vigorous wind does manage to disrupt some of the disk,
dragging it out to form a cone within a much larger bubble. This
provides a natural explanation
for the difference in distribution between the outflow cone seen in
H$\alpha$ and the more extended \mbox{X-ray} emission.
Without an early mild
wind (models B1 and B2), the disk is disrupted before the wind has easy
access to the halo, and so no obvious cone in the \mbox{X-ray} is visible.
Even when present, the conical wind does not provide appreciable
soft \mbox{X-ray} flux compared to the shocked halo (see Figs. 4, 10
and 14 in SBHL), so we would not expect to see such structures in the
{\em ROSAT} data. The radial extent of the soft \mbox{X-ray} emitting
material is much greater than that of the cones, the emission appearing
more like a figure-of-eight (Fig. 4 in SBHL) or a cylinder (Fig. 10
in SBHL). None of the models SBHL present would be seen as more conical
than cylindrical when projected along the line of sight and observed by a
real \mbox{X-ray} instrument. As SBHL did not provide projected surface brightness
plots, it is difficult to assess how limb-brightened the emission would be,
but given the brightness of the shocked halo material it should be a
detectable effect.

The luminosity varies greatly between the different models, and is not
simply proportional to the mass and energy input in the starburst. Model
A1 has a time averaged mass and energy input an order of magnitude less
than model B1, but has a soft ($0.1-2.2\keV$) luminosity of $4\times10^{41}
\ergps$ at $t=16.6$ Myr (when the starburst luminosity is $\sim10^{44}
\ergps$, very close to M82's bolometric luminosity), well above the
$0.1-2.4\keV$ wind luminosity of $\sim2\times10^{40} \ergps$ derived above
for M82.  Model B1 has a corresponding \mbox{X-ray} luminosity of only
$1.3\times10^{40} \ergps$, despite having a similar initial gas distribution to
model A1.  Given the wide range of predicted wind luminosity it would require a
deeper investigation of the available parameter space to use M82's luminosity
to constrain the allowable models.

SBHL provide ``effective'' temperatures for the gas in their ``soft''
band which corresponds well to the {\em ROSAT} band, by comparing
fluxes in two energy bands of $0.1-0.7\keV$ and $0.7-2.2\keV$, as well as
the temperature range for the gas that provides the majority of the soft
emission. In all cases the emission is very soft, typically $T\sim0.2\keV$.
The hottest model (B1) has a characteristic temperature of only $0.4\keV$.
This is still cooler than M82's wind, where the temperature varies
between $0.4-0.6\keV$. SBHL stress the lack of appreciable amounts of gas
hotter than $\sim6\times10^{6} \K$. The wind itself is much hotter, but
provides very little emission, $L_{X}\sim10^{38} \ergps$. The lower
temperatures predicted by SBHL do correspond with observations of
some galaxies for which soft \mbox{X-ray} emission can be detected.
NGC 891 has a halo with $T\sim0.3\keV$ (Bregman \& Pildis, 1994), and
Wang \etal (1995) detect soft $\sim0.25\keV$ \mbox{X-ray} emission out to
more than $8\kpc$ from the plane of NGC 4631. A survey of {\em ROSAT}
PSPC observation of nearby normal and starburst galaxies by Read \etal
(1996) shows that many starbursts have diffuse gas with temperatures
$\sim0.5\keV$, more in line with our results.

The density of the gas responsible for the emission in SBHL's
models is typically $n\sim4-8 \times 10^{-2} \pcc$. This is not inconsistent
with our results of $n_{\rm e}\sim1-4 \times 10^{-2} \eta^{-1/2} \pcc$. The
filling factor of the emitting gas in their models we can roughly
estimate as $\sim0.1$, which bring their densities close to ours.
The gas mass providing the bulk of the soft emission depends strongly
on the model used, but model B1 which
has a soft \mbox{X-ray} luminosity similar to that observed for M82's wind
has a gas mass of $10^{6} \Msol$, comparable with the total mass derived
from the PSPC of $1.3\times10^{7} \eta^{1/2}\Msol$ for reasonable values
of the filling factor.

\section{Conclusions}
We have provided a detailed spectral investigation of the extended soft
\mbox{X-ray} emission associated with the starburst galaxy M82.  Point sources
have been located and removed from the diffuse wind emission, and the effects
of contamination by the nuclear source allowed for.  The diffuse
emission was divided into a set of distinct regions as a function of distance
from the galactic plane, allowing temperature, emission measure and absorbing
column to be derived as a function of distance along the minor
axis.  The metallicity was found to be apparently low 
(0.00--$0.07 Z_{\odot}$) throughout the wind, in agreement with results
from ASCA. This work has shown the following:

\begin{itemize}
\item The observed soft \mbox{X-ray} morphology differs
significantly between
north and south, and is not well described as a conical
outflow. The emission
extends out to $\sim6\kpc$ from the plane of the galaxy. There is no
evidence for any limb brightening, as would be expected if the soft
emission came from the shock heated halo surrounding the hot wind.
\item The emission from the wind is thermal. The temperature drops slowly
along the wind, from $\sim0.6\keV$ near the nucleus,
to $\sim0.4\keV$ in the outer wind. Numerical models of
galactic winds predict the majority of the emission in the {\em ROSAT}
band to be shocked halo, with effective temperatures in the range
$0.2-0.4\keV$.
\item Since the entropy of the gas is not constant, at least in the south,
the observed emission cannot originate from a free wind itself.
Our baseline analysis is based on an assumed cylindrical geometry, but
a more divergent geometry only makes the southern wind less isentropic.
\item The emission cannot come from a static halo of gas around M82,
given that the mass required to bind a hydrostatic halo is an order of
magnitude higher than that inferred from the \hi rotational velocity in the
outer disk. 
\item For reasonable mass and energy input from the starburst, a simple
model for emission from shock heated clouds using the observed
density of the \mbox{X-ray} emitting gas to predict post-shock
temperatures is consistent with the observed temperatures.
\end{itemize}

\begin{acknowledgements}
We thank the anonymous referee for valuable comments that
have lead to the improvement of this paper. 
We acknowledge the use of the {\em Starlink} node at the University of
Birmingham and thank the developers and maintainers of the
\mbox{X-ray} analysis package ASTERIX.
DKS and IRS gratefully acknowledge PPARC funding.
This research has made use of data obtained from the UK ROSAT Data
Archive Centre at the Department of Physics and Astronomy, Leicester
University, UK, and the SIMBAD astronomical database at the CDS,
Strasbourg.
\end{acknowledgements}

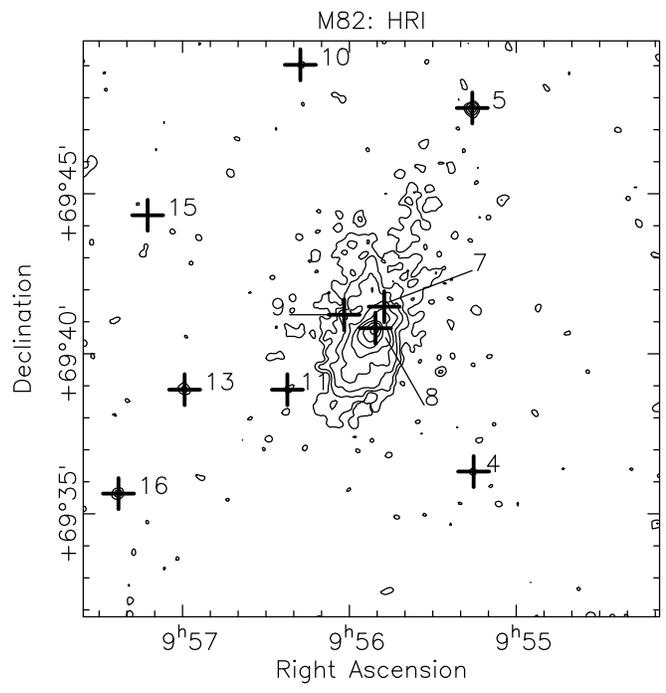
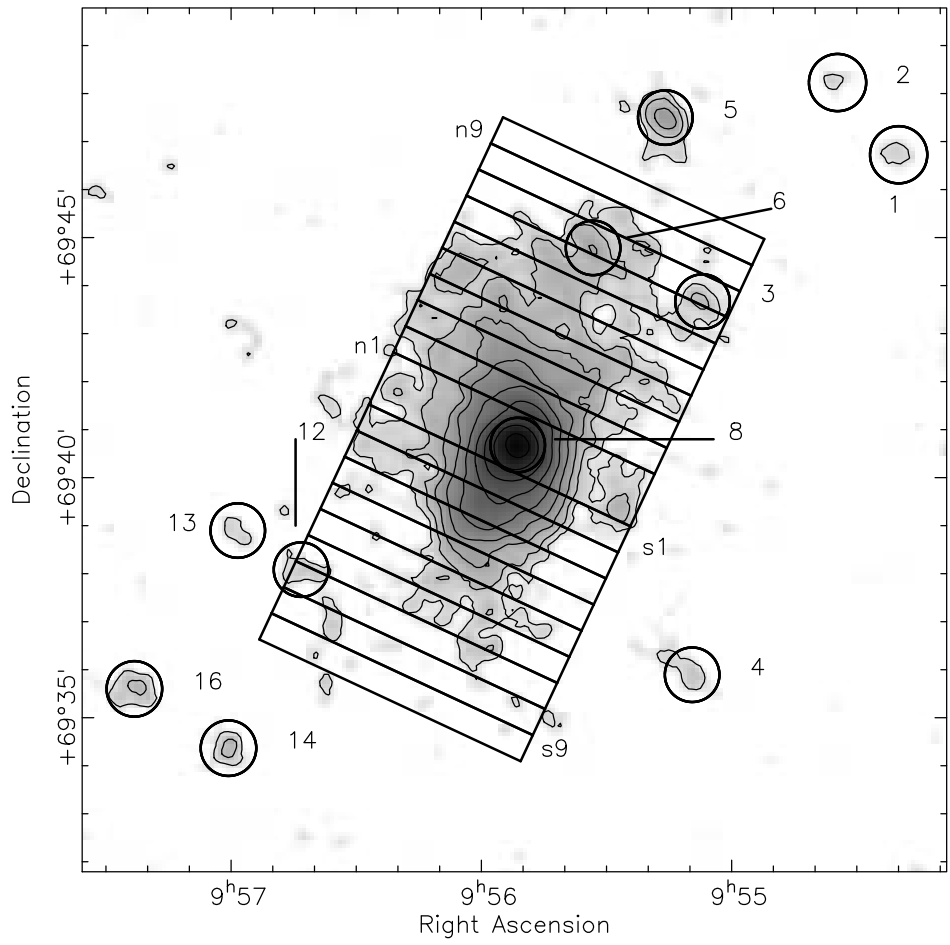
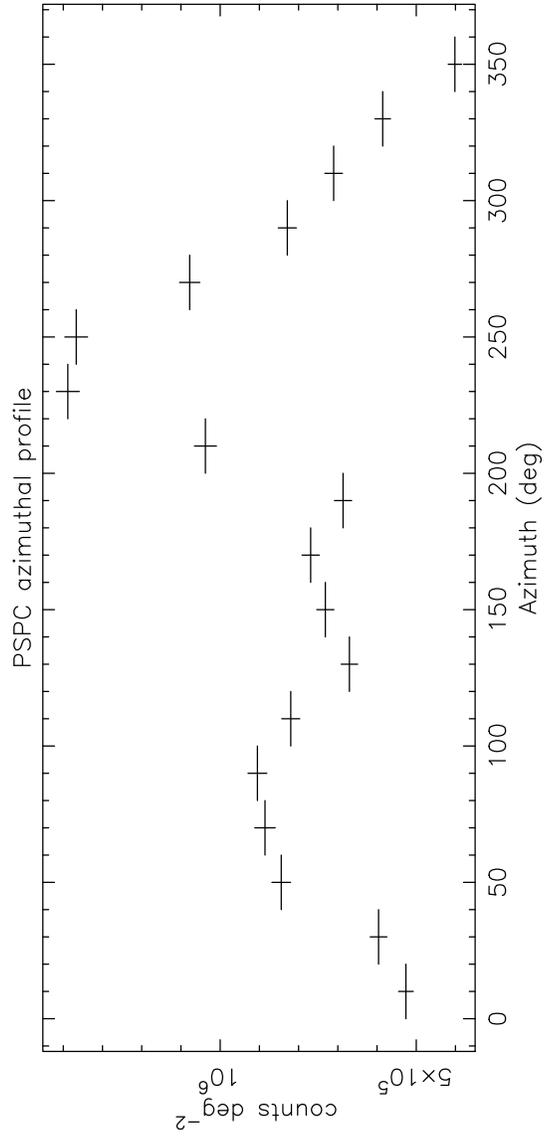
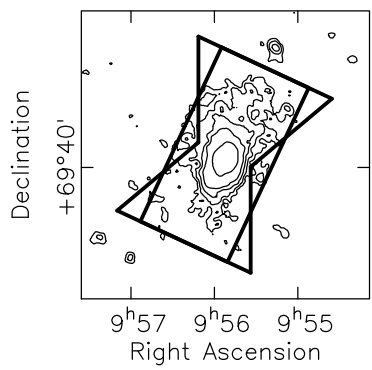
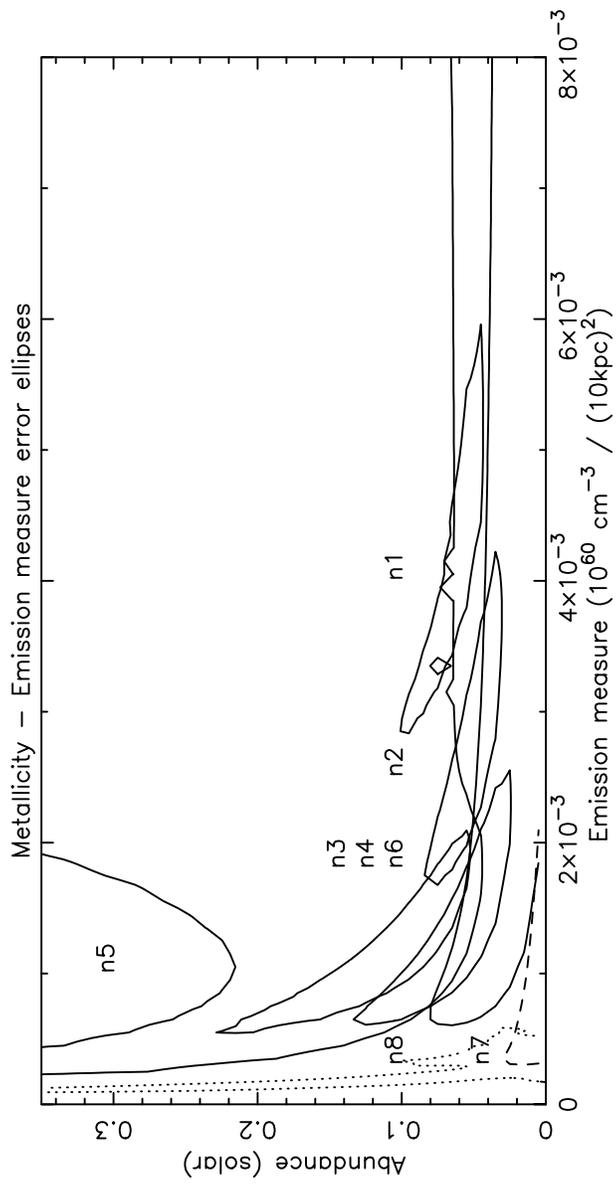
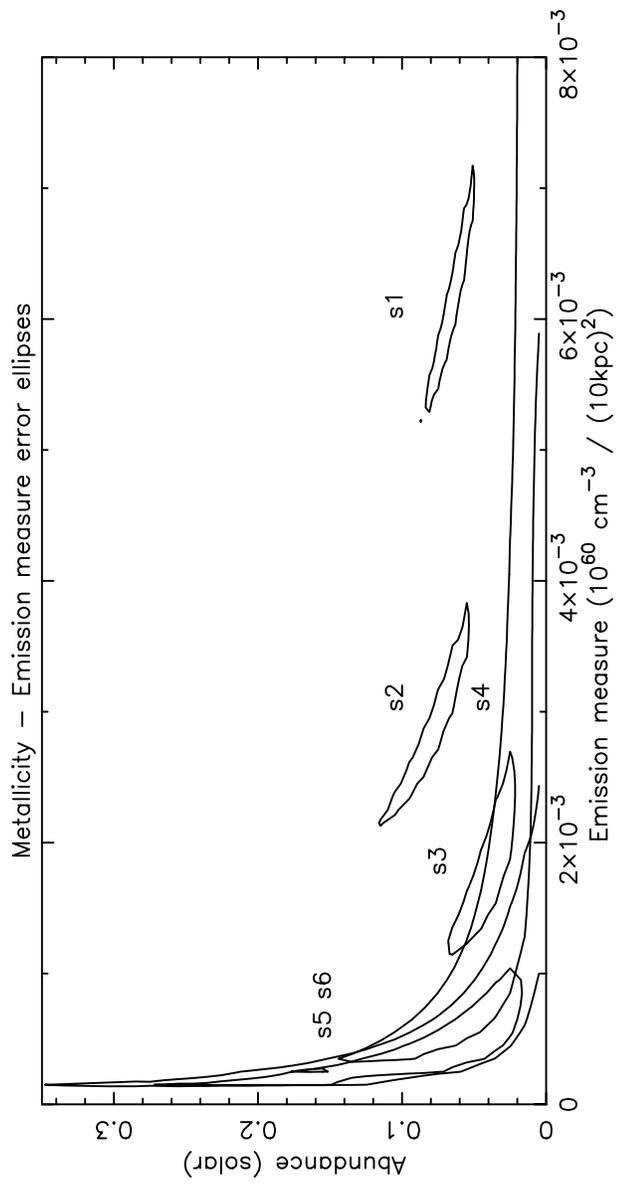
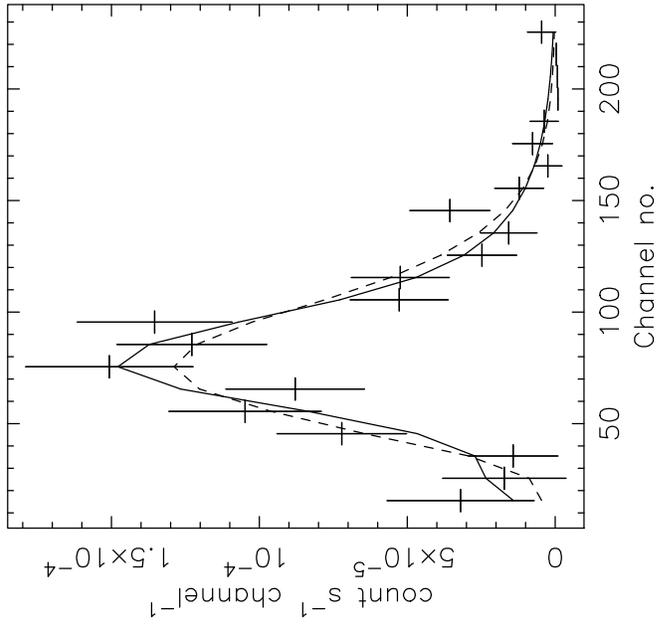
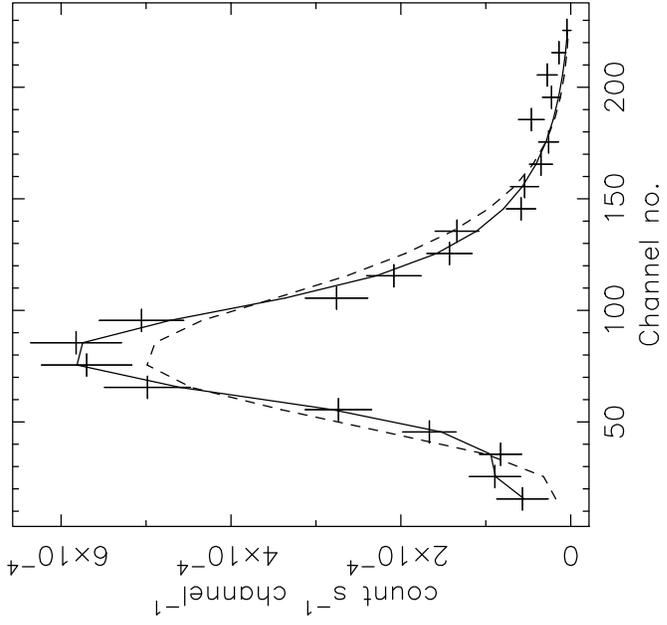
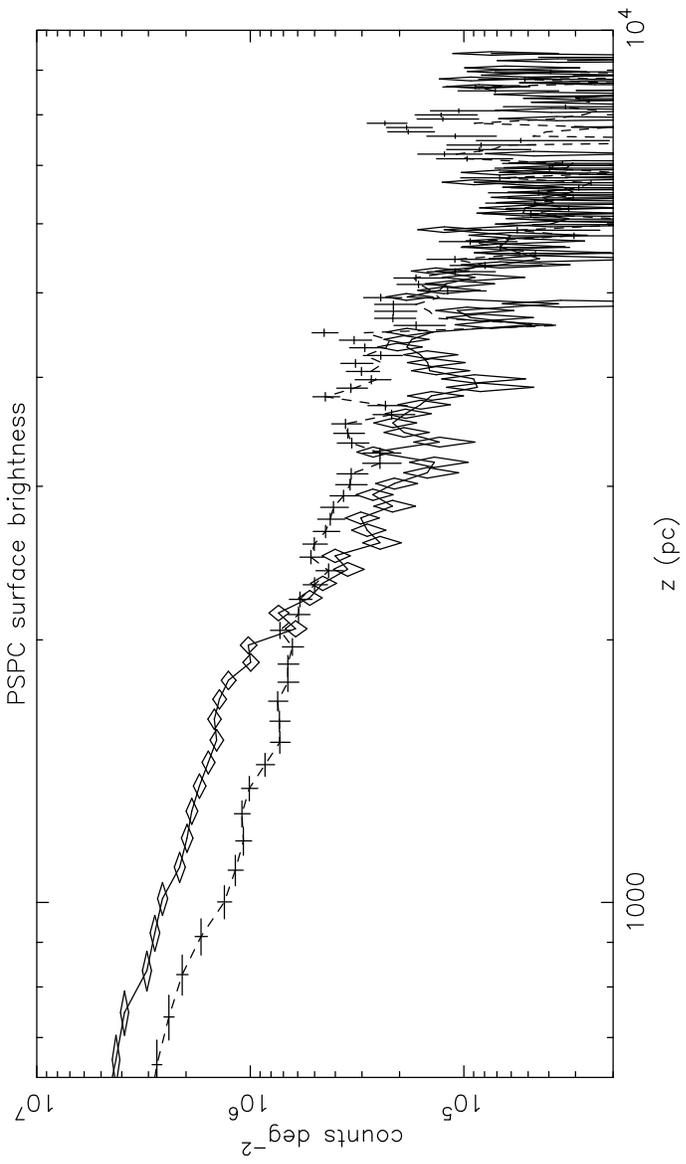
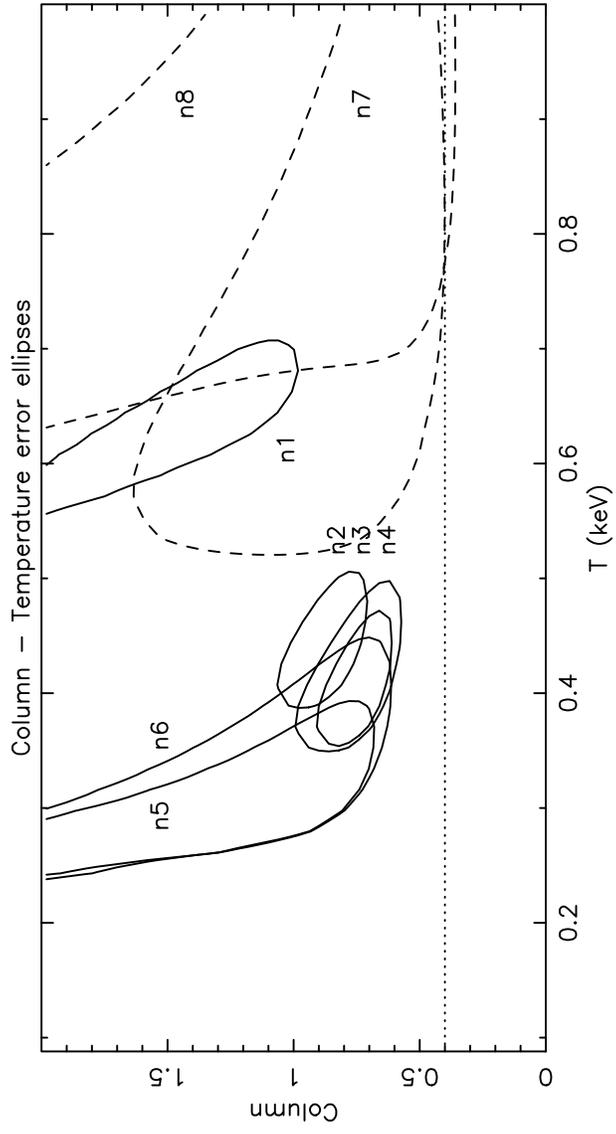
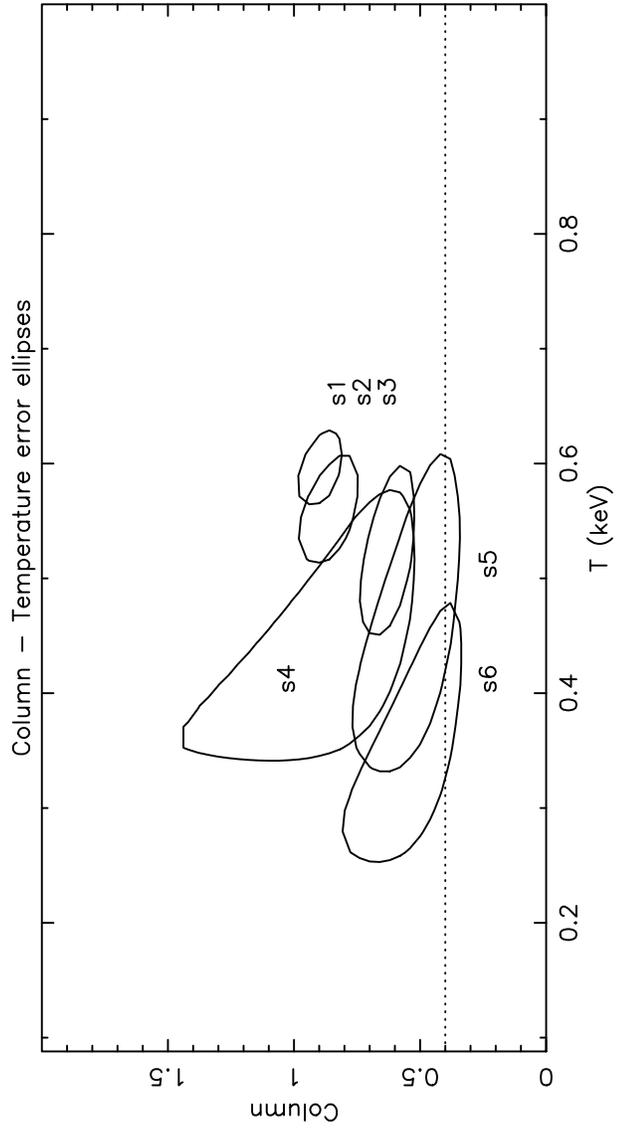
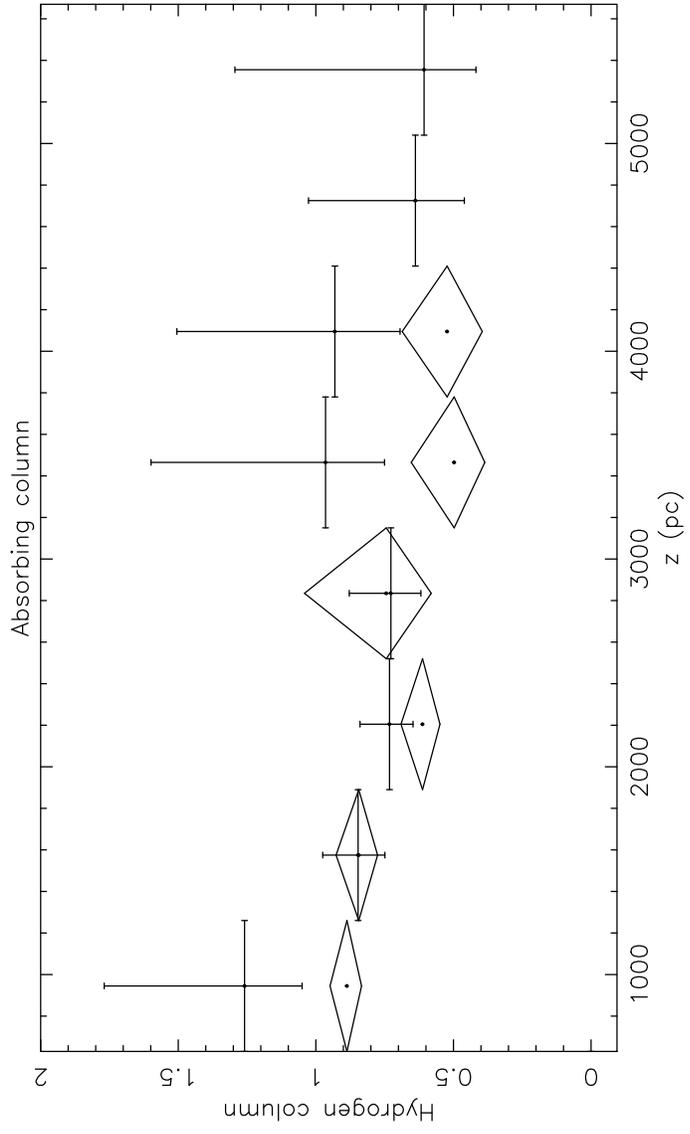
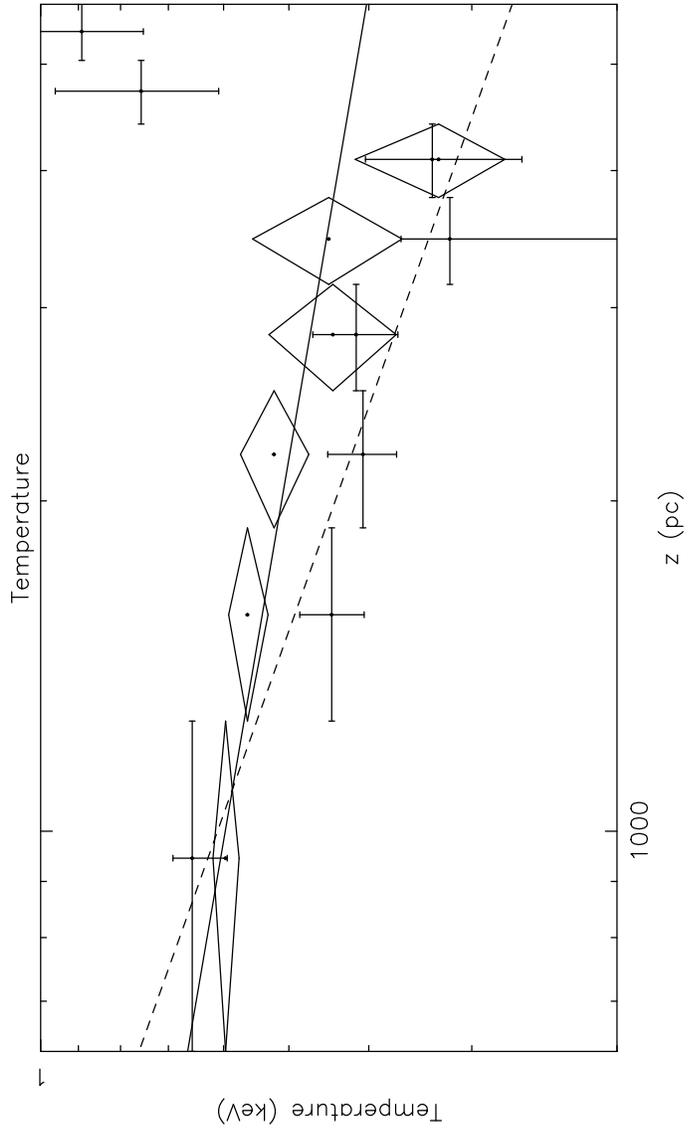
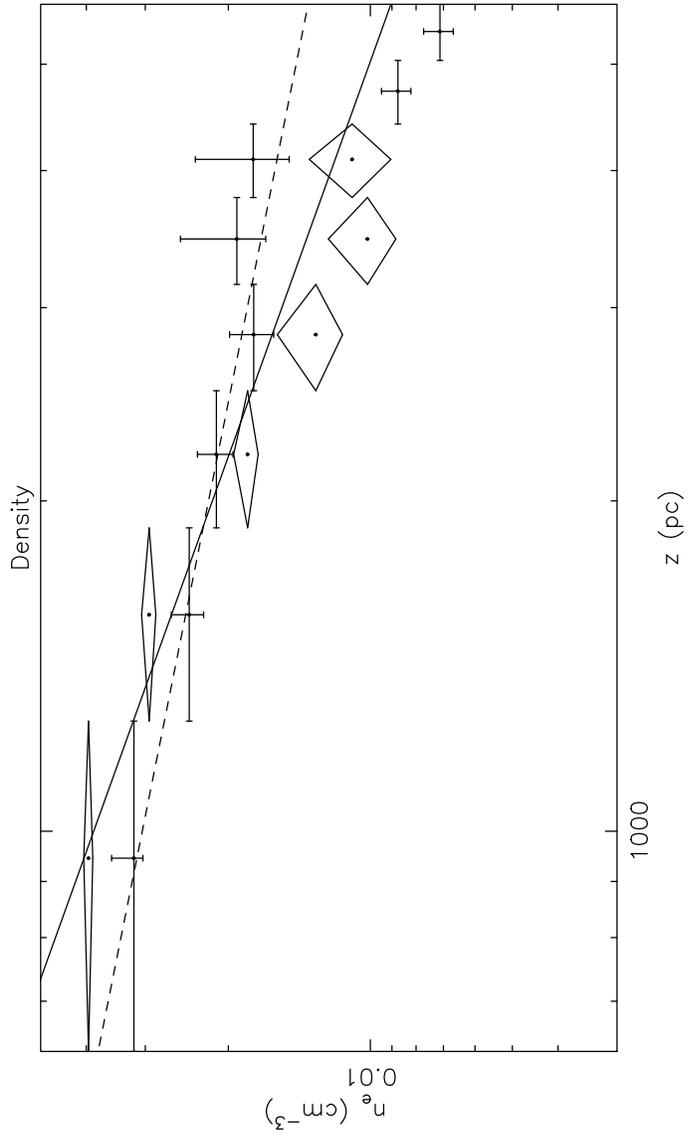
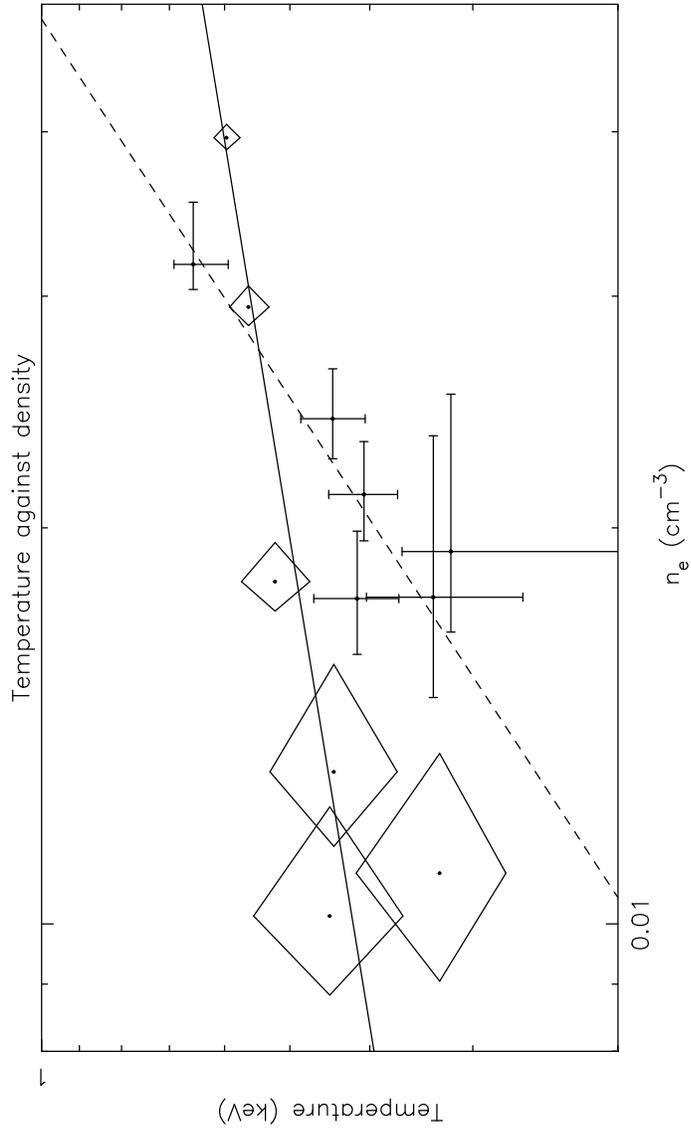
\end{document}